\documentclass[a4paper,onecolumn,11pt,unpublished]{quantumarticle}
\pdfoutput=1
\usepackage[utf8]{inputenc}
\usepackage[english]{babel}
\usepackage[T1]{fontenc}
\usepackage{amsmath}
\usepackage{hyperref}

\usepackage{tikz}

\usepackage[numbers]{natbib}
\usepackage{amssymb, amsfonts}
\usepackage{physics}
\usepackage{algorithm}
\usepackage{algpseudocode}
\usepackage{graphicx}
\usepackage{quantikz}
\usepackage{booktabs}
\usepackage{array}
\usepackage[margin=1in]{geometry}
\usepackage{amsthm}
\usepackage{makecell}
\usepackage{enumitem}
\usepackage[capitalise]{cleveref}

\newtheorem{theorem}{Theorem}
\newtheorem{lemma}{Lemma}
\newtheorem{proposition}{Proposition}
\newtheorem{corollary}{Corollary}
\theoremstyle{definition}
\newtheorem{definition}{Definition}
\theoremstyle{remark}
\newtheorem{remark}{Remark}

\DeclareMathOperator{\Hess}{Hess}
\DeclareMathOperator*{\argmin}{arg\,min}

\newcommand{\R}{\mathbb{R}}

\DeclareMathOperator{\Crit}{Crit}
\DeclareMathOperator{\codim}{codim}

\algrenewcommand\algorithmicrequire{\textbf{Input:}}
\algrenewcommand\algorithmicensure{\textbf{Output:}}

\begin{document}

\title{Double-Bracket Quantum Algorithms for Preparing Multiple Excited States}

\author{Salahuddin Abdul Rahman}
\affiliation{Automation and Control Section, Department of Electronic Systems, Aalborg University, Aalborg, Denmark}
\orcid{0009-0002-9686-8586}
\email{saabra@es.aau.dk}
\author{Özkan Karabacak}
\orcid{0000-0002-8350-193X}
\affiliation{Department of Mechatronics Engineering, Kadir Has University, Istanbul, Turkey}
\author{Rafal Wisniewski}
\affiliation{Automation and Control Section, Department of Electronic Systems, Aalborg University, Aalborg, Denmark}
\orcid{0000-0001-6719-8427}

\maketitle

\begin{abstract}
Many problems in quantum chemistry, many-body physics, and optimization require excited states of a Hamiltonian. Double-bracket quantum algorithms compile circuits recursively from evolutions under the Hamiltonian and a second operator, without a parameterized ansatz. They have been proposed for several tasks, such as diagonalization and the preparation of ground and thermal states. In this work, we extend this framework to excited states. Combining the ensemble approach with double-bracket flows, we introduce the Double-Bracket quantum algorithm for Multiple EXcited states (DB-MEX), which evolves $p+1$ orthonormal states toward low-lying eigenstates of the Hamiltonian with a single recursively compiled quantum circuit, so they remain orthonormal by construction. The circuit approximates a double-bracket flow of a weighted sum of projectors onto the evolving states, whose number and weights are design parameters. The number of states sets the highest targeted excited state, and the weights determine whether the algorithm prepares the $p+1$ lowest eigenstates, only the $p$-th excited state, or only the span of the $p+1$ lowest eigenstates. In the latter case, a post-processing step recovers the individual eigenstates from their span. We show that the double-bracket flow and, for small step sizes, its discretization converge to the target for almost all initial states, with explicit local rates given by products of weight and energy gaps. 
\end{abstract}

\section{Introduction}
\label{sec:intro}
Preparing low-lying excited states of a Hamiltonian is important in quantum chemistry~\cite{parrish2019quantum}, many-body physics~\cite{fauseweh2024quantum}, and optimization~\cite{ABDULRAHMAN2026107979}. For local Hamiltonians, estimating the ground-state energy or the energy of a low-lying excited state is QMA-hard~\cite{kitaev2002classical,bookatz2012qma}. Thus, these problems are not expected to admit efficient algorithms in general, even on quantum computers. However, worst-case hardness does not imply that every instance is hard, and practical instances often exhibit structure or admit good initial approximations.

Many quantum algorithms for excited states have been proposed. Some prepare eigenstates directly on the quantum computer, such as quantum phase estimation~\cite{abrams1999quantum} and adiabatic evolution~\cite{albash2018adiabatic}, while others extract them classically from measured matrices, such as quantum subspace expansion~\cite{mcclean2017hybrid}, quantum equation of motion~\cite{ollitrault2020quantum}, and quantum Lanczos methods~\cite{motta2020determining}. Another group reduces the problem to ground-state preparation by techniques such as deflation, spectrum folding, or shift-invert transformations~\cite{higgott2019variational,jones2019variational,tsuchimochi2023improved,11355955,millar2026imaginary}. Ensemble methods use the fact that a Hermitian operator has an orthogonal eigenbasis and that unitary evolution preserves orthogonality. They evolve several orthogonal states, typically under a shared unitary, toward the target eigenstates~\cite{parrish2019quantum,nakanishi2019subspace,yalouz2021state,rahman2024weighted,tsuchimochi2023multi,cianci2024subspace}, where the unitaries are obtained by classical optimization or from linear equations built from measured quantities. We compare these in detail in Sec.~\ref{sec:related}.

Brockett introduced double-bracket flows for matrix diagonalization and sorting~\cite{brockett1991dynamical}, and Helmke and Moore~\cite{helmke2012optimization} discuss their properties in detail. Recently, double-bracket quantum algorithms have been proposed for diagonalization~\cite{gluza2024double}, ground-state preparation~\cite{gluza2026double,robbiati2026double}, thermal-state preparation~\cite{wright2026double}, quantum signal processing~\cite{suzuki2025double}, and unstructured search~\cite{suzuki2025grover}. They compile circuits coherently from evolutions under the Hamiltonian and a second operator, without a parameterized ansatz.

In this work, we combine the ensemble approach with double-bracket flows to prepare several low-lying eigenstates with a single quantum circuit, and introduce the Double-Bracket quantum algorithm for Multiple EXcited states (DB-MEX). DB-MEX evolves a weighted sum of projectors onto $p+1$ orthonormal states, which we call the ensemble operator, under a double-bracket flow. The rank and the weights of this operator are design parameters. The rank $p+1$ sets the highest targeted excited state, and the weights determine whether DB-MEX prepares the $p+1$ lowest eigenstates, only the $p$-th excited state, or their span, which a post-processing step then resolves. Double-bracket quantum imaginary-time evolution (DB-QITE)~\cite{gluza2026double} uses a rank-one operator and targets the ground state, while double-bracket diagonalization~\cite{gluza2024double} targets the whole spectrum, even when only a few eigenstates are needed. DB-MEX instead targets exactly the eigenstates of interest by choosing the rank and the weights, and its local rates depend only on the gaps around them and on the weight gaps.

We prove that the double-bracket flow of the ensemble operator converges to its target for almost every initial ensemble, including degenerate energy levels and repeated weights, with explicit local rates given by products of weight and energy gaps. These results build on classical work on double-bracket flows~\cite{helmke2012optimization}, see also Refs.~\cite{schulte2010gradient,malvetti2024randomized} for their use in quantum information and quantum dynamics. For the discretization of the flow, we show that the weighted energy decreases at every step and that, for small step sizes, the iterates converge to the same target. We then compile the iteration into a recursive circuit whose outputs are orthonormal by construction. It has the same recursive structure as the DB-QITE circuit, with the multi-controlled phase gate replaced by a weighted phase gate, for which we give an explicit implementation.

The paper is organized as follows. Sec.~\ref{sec:prelim} introduces the notation and reviews DB-QITE, and Sec.~\ref{sec:related} discusses related work. Sec.~\ref{sec:analysis} introduces the double-bracket flow of the ensemble operator and its discretization, and states their convergence results. Sec.~\ref{sec:algorithm} compiles the iteration into a circuit and describes the post-processing, and Sec.~\ref{sec:conclusion} concludes. Proofs and circuit details are given in the appendices.

\section{Preliminaries}
\label{sec:prelim}
In this section, we introduce the notation and review the DB-QITE algorithm.

\subsection{Notation}
\label{sec:notation}

We denote the state space by $\mathcal H=\mathbb C^L$ and the group of unitary operators on $\mathcal H$ by $\mathrm{U}(L)$. The target Hamiltonian $\hat H$ acting on $\mathcal H$ has eigenvalues $E_0\le E_1\le\dots\le E_{L-1}$, counted with multiplicity. We call the distinct eigenvalues of $\hat H$ its energy levels, and a level is degenerate if it has multiplicity greater than one. We write $g_n:=E_{n+1}-E_n$, ${n=0,\dots,L-2}$, for the gaps of $\hat H$. Let $\ket{E_0},\dots,\ket{E_{L-1}}$ denote corresponding orthonormal eigenvectors defined uniquely up to phase factors if all energy levels are nondegenerate or otherwise defined up to phase factors and eigenspace-invariant unitaries. For an operator $X$ on $\mathcal H$, we write $\|X\|$ for its operator norm, the largest singular value of $X$, and $\|X\|_{HS}:=\sqrt{\Tr(X^\dagger X)}$ for its Hilbert--Schmidt norm. For a closed set $C$ of operators, we write $\operatorname{dist}(X,C):=\min_{Y\in C}\|X-Y\|_{HS}$ for the distance from $X$ to $C$.

\subsection{Double-Bracket Quantum Imaginary-Time Evolution}
\label{sec:dbqite}
The normalized imaginary-time evolution (ITE) of a state $\ket\psi$ obeys
\begin{equation} \label{eq:ite}
    \frac{d}{d\tau}\ket{\psi(\tau)}=-\big(\hat H-E(\tau)\big)\ket{\psi(\tau)},
\end{equation}
where $E(\tau)=\bra{\psi(\tau)}\hat H\ket{\psi(\tau)}$. With the density matrix $\Psi:=\ket\psi\bra\psi$, Gluza et al.~\cite{gluza2026double} observed that this equation can be written as
\begin{equation} \label{eq:ite-dbf}
    \frac{d}{d\tau}\ket{\psi(\tau)}=[\Psi(\tau),\hat H]\ket{\psi(\tau)},
\end{equation}
which, in terms of the density matrix, is equivalent to
\begin{equation} \label{eq:ite-ddbf}
    \frac{d\Psi(\tau)}{d\tau}=\big[[\Psi(\tau),\hat H],\Psi(\tau)\big].
\end{equation}
This is Brockett's double-bracket flow~\cite{brockett1991dynamical,helmke2012optimization}. The generator of the evolution $[\Psi,\hat H]$ is skew-Hermitian, so the state evolves by a state-dependent unitary.

Discretizing this continuous flow with a step size $s_k$, using a piecewise-constant Lie--Euler approximation of the state-dependent generator, yields the discrete double-bracket iteration for the state,
   \begin{equation} \label{eq:dbi-single}
       \ket{\psi_{k+1}}=e^{s_k[\Psi_k,\hat H]}\ket{\psi_k},
   \end{equation}
in which each step is an exact unitary. For suitable step sizes, this iteration converges to the ground state of $\hat H$, provided the initial state $\ket{\psi_0}$ has a nonzero overlap with the ground state (see Ref.~\cite{gluza2026double} and references therein).

Following Ref.~\cite{gluza2026double}, we distinguish three terms. ITE is the normalized non-unitary evolution~\eqref{eq:ite}. Quantum imaginary-time evolution (QITE) refers to a unitary implementation of ITE on a quantum computer. DB-QITE is a specific QITE algorithm, obtained by compiling the iteration~\eqref{eq:dbi-single} into quantum circuits. More generally, double-bracket quantum algorithms are recursive methods for compiling quantum circuits from double-bracket flows. They were proposed to obtain diagonalizing quantum circuits for a Hamiltonian~\cite{gluza2024double}, and DB-QITE applies the same construction to ITE~\cite{gluza2026double}.

In Sec.~\ref{sec:analysis}, we replace the rank-one projector $\Psi$ with a rank-$(p+1)$ operator $\Omega$ and show how a single double-bracket evolution prepares several excited states. For $p>0$, this evolution is in general no longer an ITE of the individual states, so the resulting method is a double-bracket quantum algorithm but not a QITE algorithm in the above sense. In Sec.~\ref{sec:algorithm}, we compile its discretization into the quantum algorithm DB-MEX.

\section{Related Work}
\label{sec:related}
A common approach to excited-state preparation is to convert it into a ground-state problem, using deflation~\cite{saad2011numerical}, the folded-spectrum method~\cite{wang1994solving}, or shift-invert techniques~\cite{SciPostPhys}. These techniques have been adapted to quantum algorithms, including variational quantum eigensolvers (VQEs)~\cite{higgott2019variational,mcclean2016theory} and feedback-based quantum algorithms (FQAs)~\cite{ABDULRAHMAN2026107979,11355955}. The deflation approach modifies the Hamiltonian to $\hat H+\sum_{q=0}^{p-1}\alpha_q\ketbra*{\tilde e_q}$, where $\{\tilde e_q\}$ are approximations of the $p$ lowest eigenstates and the penalties $\{\alpha_q\}$ are hyperparameters that must exceed the relevant energy gaps. The ground state of this Hamiltonian then approximates the $p$-th excited state of $\hat H$. Jones et al.~\cite{jones2019variational} adapted this construction to variational ITE, preparing the eigenstates sequentially, starting from the lowest one. Deflation requires careful tuning of the penalties and accurate approximations of the lower eigenstates, since their errors propagate to the higher ones.

The folded-spectrum method instead replaces $\hat H$ with $(\hat H-\lambda)^2$, whose ground state is the eigenstate of $\hat H$ with energy closest to $\lambda$. Tsuchimochi et al.~\cite{tsuchimochi2023improved} adapted this construction to QITE and found that it can slow convergence. This method requires an estimate of the target energy to choose $\lambda$ and access to $\hat H^2$, which complicates implementation. The shift-invert technique uses $(\hat H-\delta)^{-1}$, which turns the eigenstate with the largest energy below $\delta$ into the ground state. It has been adapted to ITE and demonstrated with classical tensor-network simulations~\cite{millar2026imaginary}. Like the folded-spectrum method, it requires an energy estimate to place $\delta$ appropriately in the spectrum, and it is difficult to realize on a quantum computer.

Another ITE-based approach is the quantum Lanczos method~\cite{motta2020determining,yeter2020practical}, which diagonalizes $\hat H$ in a Krylov subspace spanned by imaginary-time-evolved states. This requires measuring the Hamiltonian and overlap matrices in that subspace.

A different approach is to evolve several orthogonal states with one shared unitary. Since a unitary preserves orthogonality, minimizing the ensemble-weighted energy drives the states toward the lowest eigenstates, as follows from the ensemble variational principle~\cite{gross1988rayleigh,ding2024ground}. This approach has been adapted to VQE~\cite{parrish2019quantum,nakanishi2019subspace,yalouz2021state} and contracted quantum eigensolvers~\cite{benavides2024quantum}. In our earlier work, we applied it to FQAs and introduced the weighted feedback-based quantum algorithm for excited states (WFQAE), which builds the circuit layer by layer from measured quantities~\cite{rahman2024weighted}.

Two methods evolve orthogonal ensembles within QITE, namely model-space QITE (MSQITE)~\cite{tsuchimochi2023multi} and subspace-search QITE (SSQITE)~\cite{cianci2024subspace}. MSQITE evolves each state of an orthogonal model space with its own unitary that approximates ITE, restores orthogonality at each step by an orthonormalization procedure, and obtains the excited states by diagonalizing the Hamiltonian in the model space at the end. A state-averaged variant, in which all states share one unitary, was also considered, and its accuracy was found to be limited by the restricted form of that unitary. SSQITE applies variational ITE to an orthogonal ensemble and orders the states by assigning larger step sizes to lower-energy states, while checking orthogonality during the evolution. In both methods, the non-unitary imaginary-time step is approximated by a unitary whose parameters are obtained from linear equations built from measured quantities. Smart and Narang~\cite{smart2024many} treat the same ensemble problem as a direct Riemannian optimization over the Stiefel and Grassmann manifolds, and estimate the gradient by tomography at every step. DB-MEX instead compiles a discretized double-bracket flow into a quantum circuit built from evolutions under $\hat H$ and the ensemble operator, without fitting, tomography, or an ansatz. Orthogonality then holds by construction, and the same circuit maps each input state that carries its own weight to a prescribed eigenstate of $\hat H$.

Double-bracket diagonalization~\cite{gluza2024double} was designed to diagonalize $\hat H$. When applied to eigenstate preparation, it generates a single shared unitary that approximately maps computational basis states to the corresponding eigenstates of $\hat H$. However, it achieves this by targeting the full spectrum, even when only a few low-lying eigenstates are needed.

Compared with the methods above, DB-MEX has the following properties.
\begin{itemize}[noitemsep]
\item The circuit is compiled recursively from evolutions under $\hat H$ and the ensemble operator, without a parameterized ansatz, classical optimization, or fitting, so its accuracy is not limited by ansatz expressivity.
\item The output states are orthonormal by construction, so DB-MEX needs no penalty terms, orthonormalization, orthogonality checks, or overlap measurements.
\item Unlike the folded-spectrum and shift-invert methods, DB-MEX needs no estimates of the target energies, since the number of states in the ensemble sets the target. Unlike deflation, it needs no penalties tuned to the energy gaps and no fixed approximations of the lower eigenstates, whose errors would carry over to the excited states. The lower states instead evolve together with the target.
\item The flow and, for small step sizes, its discretization converge to the target for almost every initial ensemble, with explicit local rates, including for degenerate energy levels and repeated weights.
\item Unlike double-bracket diagonalization, DB-MEX targets only the eigenstates of interest, and its local rates depend only on the gaps around them and on the weight gaps.
\item A single run of the recursion compiles one circuit, which then prepares any targeted state when applied to the corresponding input. The shared circuit also simplifies measuring quantities such as transition amplitudes between the states, which can be obtained from expectation values without an ancilla, as in subspace-search VQE~\cite{nakanishi2019subspace}.
\end{itemize}
These properties come at the cost of a circuit depth that grows exponentially with the number of steps, as discussed in Sec.~\ref{sec:circuit}.

\section{Double-Bracket Flows for Excited States}
\label{sec:analysis}
We now introduce the double-bracket flow on which DB-MEX is built. It acts on the ensemble operator $\Omega$, a weighted sum of projectors onto $p+1$ orthonormal states, whose rank and weights are the design parameters. Sec.~\ref{sec:setting} defines the double-bracket flow on a set of ensemble operators and provides convergence guarantees, which hold for arbitrary multiplicities in the spectra of $\hat H$ and $\Omega$. Sec.~\ref{sec:LinearDynamics} constructs time-varying unitary dynamics on quantum states induced by the double-bracket flow. Sec.~\ref{sec:weights} shows how the weights select the target, and Sec.~\ref{sec:rates} evaluates the convergence rates for each choice. Sec.~\ref{sec:iteration} discretizes the flow into an iteration that updates all states with the same unitary and gives its convergence guarantees. We give the proofs in the appendices. In Sec.~\ref{sec:algorithm}, we present the DB-MEX algorithm built on these results, including the post-processing that resolves the eigenstates spanned by states that share a weight.

\subsection{Double-Bracket Flow on Ensemble Operators}
\label{sec:setting}
For a given set of weights $w_0\ge w_1\ge\dots\ge w_p>0$, we define the set of ensemble operators
\begin{equation} \label{eq:omega}
    M(\bar\Omega)=\{ U \bar\Omega U^\dagger : U \in \mathrm{U}(L)\},\quad\bar\Omega=\sum_{q=0}^p w_q\ket*{\phi^{(q)}}\bra*{\phi^{(q)}},
\end{equation}
where $\ket*{\phi^{(q)}}$, $q=0,\dots,p$, which we call ensemble states, are some arbitrary orthonormal vectors. The weights are the nonzero eigenvalues of the Hermitian operators in $M(\bar\Omega)$ and are the design parameters of our approach. Note that the weights need not sum to one, since $\bar\Omega$ is not prepared as a quantum state. Rescaling all weights only rescales the time of the designed dynamics.

Let us consider Brockett's double-bracket flow~\cite{helmke2012optimization} on $M(\bar\Omega)$:
\begin{equation} \label{eq:dbf}
    \frac{d\Omega(\tau)}{d\tau}
    =\comm{\comm{\Omega(\tau)}{\hat H}}{\Omega(\tau)}.
\end{equation}
Note that the flow changes $\Omega$ only by unitary conjugation, so its spectrum, and hence the weights, stay fixed during the evolution. It is known that the flow~\eqref{eq:dbf} is the gradient flow of the weighted energy $V(\Omega)=\Tr(\Omega\hat H)$ on $M(\bar\Omega)$. The minimum of $V$ is $V_*=\sum_{q=0}^pw_qE_q$, which is attained at the ensembles
\begin{equation} \label{eq:MOmega}
\Omega_*=\sum_{q=0}^pw_q\ket{E_q}\bra{E_q},
\end{equation}
where $\{\ket{E_q}\}$ are the orthonormal eigenvectors of $\hat H$ as uniquely defined in Section~\ref{sec:prelim} up to phase factors and eigenspace-invariant unitaries. Note that, in the above expression, the largest weights are associated with the lowest eigenstates. Let us define the set of all such ensembles as
\begin{equation} \label{eq:Cstar}
    C_*:=\argmin_{\Omega\in M(\bar\Omega)}\Tr{\Omega \hat{H}},
\end{equation}
which is a singleton unless a pair of degenerate eigenstates is associated with different weights, counting $w_n=0$ for $n>p$. The following theorem is a generalization of \cite[Theorem~1.5]{helmke2012optimization} to the case of Hermitian matrices with possibly degenerate eigenvalues:

\begin{theorem}[Convergence of the flow]\label{thm:flow}
For a Hermitian matrix $\hat H$, consider the flow \eqref{eq:dbf} on $M(\bar\Omega)$ defined in \eqref{eq:omega}  and let $C_*$ be defined as in \eqref{eq:Cstar}. Then,
\begin{enumerate}

\item for every initial ensemble $\Omega(0)=\Omega_0$, $\Omega(\tau)$ converges to an ensemble that commutes with $\hat H$;

\item for almost all initial ensembles $\Omega(0)=\Omega_0$, $\lim_{\tau\to\infty}\Omega(\tau)$ lies in $C_*$. 
\end{enumerate}
\end{theorem}
The proof is given in Appendix~\ref{app:morse-bott}. Item 2 excludes a set of initial ensembles of measure zero (see Appendix~\ref{app:morse-bott} for its structure) that converge to other ensembles commuting with $\hat H$. These limits are critical points of $V$ other than minima, and almost every small perturbation of such an initial ensemble leads to $C_*$ instead. For example, the excluded set contains every ensemble in which the eigenspace of a targeted level $E_m$, $m\le p$, is orthogonal to all initial ensemble states. The states then remain orthogonal to it, so this level is never reached. For $p=0$, this reduces to the condition of ITE, namely that the state converges to the ground eigenspace if and only if it has a nonzero overlap with that eigenspace, that is, $\braket*{E_0}{\phi(0)}\neq0$ when the ground state is unique.

\subsection{Obtaining Linear Dynamics on State Vectors}
\label{sec:LinearDynamics}
We now obtain a nonautonomous linear matrix differential equation from the double-bracket flow and show that its solutions can be reduced to solutions of a set of uncoupled nonautonomous linear vector differential equations, which we discretize in Sec.~\ref{sec:iteration} and compile into a quantum circuit in Sec.~\ref{sec:algorithm}.

For a particular solution $\tilde\Omega(\tau)$ of \eqref{eq:dbf}, let us define 
\begin{equation} \label{eq:ParticularSolution}
    A(\tau):=\comm{\tilde\Omega(\tau)}{\hat H},
\end{equation}
which is skew-Hermitian. The solution of the double-bracket flow \eqref{eq:dbf} for the initial ensemble $\tilde\Omega(0)$ is also the solution to the following nonautonomous linear matrix differential equation 
\begin{equation} \label{eq:MatrixDif}
    \frac{d\Omega(\tau)}{d\tau}
    =\comm{A(\tau)}{\Omega(\tau)},\quad \Omega(0)=\tilde\Omega(0),
\end{equation}
which is isospectral, since $A(\tau)$ is skew-Hermitian. By the standard representation of isospectral flows \cite[Section IV.3.2]{hairer2006}, the solution of \eqref{eq:MatrixDif} admits the spectral representation
\begin{equation} \label{eq:SpectralRepresentation}
\Omega(\tau)=\sum_{q=0}^{p}w_q \ket*{\phi^{(q)}(\tau)}\bra*{\phi^{(q)}(\tau)},
\end{equation}
where the orthonormal eigenvectors $\ket*{\phi^{(q)}(\tau)}$ satisfy the following nonautonomous linear differential equation
\begin{equation} \label{eq:ite-dbf-nonautonomous}
    \frac{d}{d\tau}\ket{\phi(\tau)}=A(\tau)\ket{\phi(\tau)}, \quad \ket*{\phi(0)}=\ket*{\phi^{(q)}(0)}.
\end{equation}

\subsection{Weights and Target States}
\label{sec:weights}
Theorem~\ref{thm:flow} shows that, for almost every initial ensemble $\Omega(0)=\sum_q w_q\ket*{\phi^{(q)}(0)}\bra*{\phi^{(q)}(0)}$, the ensemble trajectory $\Omega(\tau)=\sum_q w_q\ket*{\phi^{(q)}(\tau)}\bra*{\phi^{(q)}(\tau)}$ converges to $C_*$, where $\ket*{\phi^{(q)}(\tau)}$ satisfy \eqref{eq:ite-dbf-nonautonomous} as shown in the previous section. Note that each ensemble state is an eigenvector of $\Omega$ with eigenvalue $w_q$. If all energy levels of $\hat H$ are nondegenerate, an ensemble state $\ket*{\phi^{(q)}(\tau)}$ whose weight is not shared with another ensemble state converges to an eigenstate $\ket*{E_q}$ of $\hat H$, while states with a shared weight converge only to the span of the corresponding eigenstates. If the $p+2$ lowest eigenvalues of $\hat H$ are distinct, this gives the following.

\begin{corollary}\label{cor:states}
Let $\hat H$ be a Hermitian matrix with first $p+2$ eigenvalues satisfying $E_0<\dots<E_p<E_{p+1}$ and $\Omega(0)=\sum w_q\ket*{\phi^{(q)}(0)}\bra*{\phi^{(q)}(0)}$ be a generic (in the sense of Theorem~\ref{thm:flow}) initial ensemble. The trajectories of the ensemble states $\ket*{\phi^{(q)}(\tau)}$ under the dynamics \eqref{eq:ite-dbf-nonautonomous} satisfy the following conditions up to multiplication by a phase factor:
\begin{enumerate}
\item (distinct weights) If  $w_0>w_1>\dots>w_p$, $\lim_{\tau\to\infty}\ket*{\phi^{(q)}(\tau)}=\ket{E_q}$ for every $q$;
\item (targeted weights) If $w_0=\dots=w_{p-1}>w_p$, $\lim_{\tau\to\infty}\ket*{\phi^{(p)}(\tau)}=\ket{E_p}$, and the other ensemble states converge to the span of $\ket{E_0},\dots,\ket{E_{p-1}}$;
\item (equal weights) If $w_0=\dots=w_p$, all ensemble states converge to the span of $\ket{E_0},\dots,\ket{E_p}$ (the post-processing in Sec.~\ref{sec:postprocessing} can be used to recover the eigenstates).
\end{enumerate}
\end{corollary}

If one of the targeted energy levels is degenerate, the ensemble states that converge to its eigenspace are eigenstates of $\hat H$ with that energy, but the particular eigenstates depend on the initial ensemble.

\subsection{Convergence Rates} \label{sec:rates}
 
As shown in Appendix~\ref{app:morse-bott}, the convergence is exponential near $C_*$. We now specify the local rate of convergence and evaluate it for the different choices of the weights. Recall that $w_0\ge w_1\ge\dots\ge w_p>0$ and $g_n=E_{n+1}-E_n$, and set $w_n:=0$ for $n>p$. We define
\begin{equation}\label{eq:rate}
    \mu_*:=\min\big\{(w_m-w_n)(E_n-E_m):\ w_m>w_n,\ E_m<E_n\big\},
\end{equation}
the smallest product of a weight gap and an energy gap.

\begin{theorem}[Local rate of convergence to $C_*$]
\label{thm:local-rate}
For every $0<c<2\mu_*$, there are a neighborhood $\mathcal U$ of $C_*$ and a constant $\gamma>0$ such that, if $\Omega(0)=\Omega_0\in\mathcal U$, then for all $\tau\ge0$,
\begin{enumerate}
    \item $V(\Omega(\tau))-V_*\le e^{-c\tau}\big(V(\Omega_0)-V_*\big)$,
    \item $\Omega(\tau)$ converges to a point $\Omega_\infty\in C_*$, and $\|\Omega(\tau)-\Omega_\infty\|_{HS}\le\gamma\,e^{-c\tau/2}\|\Omega_0-\Omega_\infty\|_{HS}$.
\end{enumerate}
\end{theorem}
The theorem follows from Theorem~\ref{thm:local-rate-general} in Appendix~\ref{app:morse-bott}.

If $E_0<\dots<E_{p+1}$, as in Corollary~\ref{cor:states}, the minimum in~\eqref{eq:rate} is attained at adjacent levels, and
\begin{equation}\label{eq:rate-adjacent}
    \mu_*=\min\big\{(w_q-w_{q+1})\,g_q:\ 0\le q\le p,\ w_q>w_{q+1}\big\}.
\end{equation}
Table~\ref{tab:regimes} lists this rate for the three choices of the weights. Since $\mu_*$ scales linearly with the weights, we compare the choices at a fixed largest weight $w_0$. We now discuss how to choose the weights in practice, depending on whether gap estimates are available.

For equal weights we have $\mu_*=w_0g_p$, which depends only on the gap $g_p=E_{p+1}-E_p$ between the highest targeted eigenstate and the next eigenstate, and not on the gaps between the targeted eigenstates.

For distinct weights, all gaps $g_0,\dots,g_p$ enter. Without gap estimates, the evenly spaced weights
\begin{equation*}
    w_q=w_0\bigl(1-q/(p+1)\bigr)
\end{equation*}
maximize the smallest weight gap and give 
\begin{equation*}
    \mu_*=w_0\min_{q\le p}g_q/(p+1),
\end{equation*}
which decreases with the number of targeted states. With gap estimates, choosing ${w_q-w_{q+1}}$ proportional to $1/g_q$ balances the terms of~\eqref{eq:rate-adjacent} and gives the largest local rate among distinct weights,
\begin{equation*}
    \mu_*=w_0\Big(\sum_{q=0}^{p}\frac{1}{g_q}\Big)^{-1}.
\end{equation*}

For targeted weights, $\mu_*=\min\{(w_0-w)g_{p-1},\,wg_p\}$, so only the two gaps around the $p$-th excited state enter. With gap estimates, balancing the two terms gives
\begin{equation*}
    \mu_*=w_0 \frac{g_{p-1}g_p}{g_{p-1}+g_p},
\end{equation*}
attained at $w=w_0g_{p-1}/(g_{p-1}+g_p)$.

These choices maximize only the local rate. Far from $C_*$, the trajectory can pass near saddle points of $V$, and the time spent there also depends on the weights. Hence, choosing weights that minimize total convergence time remains an open problem.

\begin{table}[H]
\centering
\small
\renewcommand{\arraystretch}{1.3}
\begin{tabular}{@{}lll@{}}
\toprule
Weights & Limit of the states & Rate $\mu_*$ \\
\midrule
\makecell[l]{Distinct:\\ $w_0>w_1>\dots>w_p>0$} &
  $\ket*{\phi^{(q)}}\to\ket{E_q}$ for all $q$ &
  $\min_{q\le p}(w_q-w_{q+1})g_q$ \\
\addlinespace
\makecell[l]{Targeted:\\ $w_0=\dots=w_{p-1}$,\\ $0<w_p=w<w_0$} &
  \makecell[l]{$\ket*{\phi^{(p)}}\to\ket{E_p}$,\\
               others $\to\operatorname{span}\{\ket{E_0},\dots,\ket{E_{p-1}}\}$} &
  $\min\big\{(w_0-w)g_{p-1},\,wg_p\big\}$ \\
\addlinespace
\makecell[l]{Equal:\\ $w_0=\dots=w_p$} &
  all $\to\operatorname{span}\{\ket{E_0},\dots,\ket{E_p}\}$ &
  $w_0g_p$ \\
\bottomrule
\end{tabular}
\caption{Choices of the weights when the $p+2$ lowest eigenvalues of $\hat H$ are distinct, $E_0<\dots<E_{p+1}$. Depending on the weights, each state converges, up to a phase, to an eigenstate of $\hat H$, or the states converge to the span of several eigenstates. The limits hold for almost every initial ensemble.}
\label{tab:regimes}
\end{table}

\subsection{Discrete Iteration}
\label{sec:iteration}

We discretize the flow~\eqref{eq:dbf} starting from the initial ensemble $\Omega_0:=\sum_{q=0}^p w_q\ket*{\phi^{(q)}_0}\bra*{\phi^{(q)}_0}$  with the Lie--Euler method as follows: 
\begin{equation} \label{eq:ensemb-update}
    \Omega_{k+1} = e^{s_k \comm{\Omega_k}{\hat{H}}} \Omega_k\, e^{-s_k \comm{\Omega_k}{\hat{H}}}.
\end{equation}
By the same argument as in Section~\ref{sec:LinearDynamics}, this induces the following iteration on $p+1$ orthonormal vectors starting from $\{\ket*{\phi^{(q)}_0}\}_{q=0,\dots,p}$:
\begin{equation}
    \ket*{\phi_{k+1}^{(q)}}=e^{s_kW_k}\ket*{\phi_k^{(q)}},
    \label{eq:ket-update}
\end{equation}
where
\begin{equation} \label{eq:Wk}
W_{k}:=\comm{\Omega_{k}}{\hat H}.
\end{equation} 
The ensemble operator $\Omega_{k+1}$ is then obtained from \eqref{eq:ensemb-update} and we keep $W_{k+1}$ fixed during the step of size $s_{k+1}$. All ensemble states are updated by the same unitary, so they remain orthonormal at every step. The generator $W_k$ is computed from the iterated ensemble states, while $A(\tau)$ in~\eqref{eq:ParticularSolution} is computed from the exact flow. The two generally differ at corresponding times, and the iterates deviate from the flow by the discretization error. The following proposition shows that the iteration nevertheless does not increase the weighted energy.

\begin{proposition}[Descent of the iteration]\label{prop:descent}
If $0<s_k<1/(2w_0\|\hat H\|)$, then
\begin{equation} \label{eq:descent}
    V(\Omega_{k+1})\le V(\Omega_k)-s_k\big(1-2s_kw_0\|\hat H\|\big)\big\|\comm{\Omega_k}{\hat H}\big\|_{HS}^2 .
\end{equation}
In particular, the weighted energy is non-increasing along the iteration and decreases strictly except on the critical sets of $V$.
\end{proposition}
The proof is given in Appendix~\ref{app:descent}. For step sizes satisfying this condition, the fixed points of the iteration are therefore exactly the ensembles that commute with $\hat H$, as for the flow.

Noticing that, for a small constant time step $s$, the trajectory of the discrete system stays near that of the flow on finite time intervals, we have the following corollary:
\begin{corollary}[Convergence of the iteration]\label{cor:iteration}
For almost every initial ensemble $\Omega_0$, there is $s_0:=s_0(\Omega_0)\in\big(0,1/(2w_0\|\hat H\|)\big]$ such that, for every constant step size $0<s<s_0$, the iterates converge to $C_*$, that is, $\operatorname{dist}(\Omega_k,C_*)\to0$.
\end{corollary}
A proof sketch of Corollary~\ref{cor:iteration} is given in Appendix~\ref{app:descent}. The ensemble states of the iteration converge as in Corollary~\ref{cor:states}, since the eigenstates they approach depend only on the limit of $\Omega_k$, which lies in $C_*$ as for the flow.

Near $C_*$, the descent bound also gives a local rate for the iteration.
\begin{proposition}[Local rate of the iteration]\label{prop:local-rate}
Let $0<s<1/(2w_0\|\hat H\|)$ be constant and $0<c<2\mu_*$. There is $\varepsilon>0$ such that, if $V(\Omega_{k_0})-V_*<\varepsilon$ for some $k_0$, then for all $k\ge k_0$,
\begin{equation*}
    V(\Omega_{k+1})-V_*\le\big(1-c\,s(1-2sw_0\|\hat H\|)\big)\big(V(\Omega_k)-V_*\big).
\end{equation*}
\end{proposition}
The proof is given in Appendix~\ref{app:descent}. Since the factor in this bound is approximately $e^{-cs}$ for small $s$, the iteration inherits the rates of Theorem~\ref{thm:local-rate}, with $\tau$ replaced by $ks$. An argument similar to the one in the proof of Theorem~\ref{thm:local-rate-general} (3) can be used to show that the iterates also converge geometrically to a point of $C_*$.

\section{DB-MEX as a Quantum Algorithm}
\label{sec:algorithm}
DB-MEX implements the iteration~\eqref{eq:ket-update} of Sec.~\ref{sec:iteration} on a quantum computer. Sec.~\ref{sec:circuit} compiles each step into a quantum circuit, recursively and approximately, using a group-commutator formula. Sec.~\ref{sec:postprocessing} describes the post-processing that resolves the eigenstates spanned by states that share a weight.

\subsection{Quantum Circuit Synthesis}
\label{sec:circuit}
Let $\{\ket{x_q}\}_{q=0}^p$ be distinct computational basis states, and let $U_0$ be either the identity or a warm-start circuit. The initial states are $\ket*{\phi^{(q)}_0}:=U_0\ket{x_q}$. Since all initial states share $U_0$, a useful warm start is a single unitary that maps every input $\ket{x_q}$ close to its target, such as the shared circuit produced by subspace-search VQE~\cite{nakanishi2019subspace}. We define the operator
\begin{equation}  \label{eq:D}
    D:=\sum_{q=0}^p w_q\ket{x_q}\bra{x_q},
\end{equation}
which is diagonal in the computational basis and has the same weights as $\bar\Omega$, so that $\bar\Omega=U_0DU_0^\dagger$.

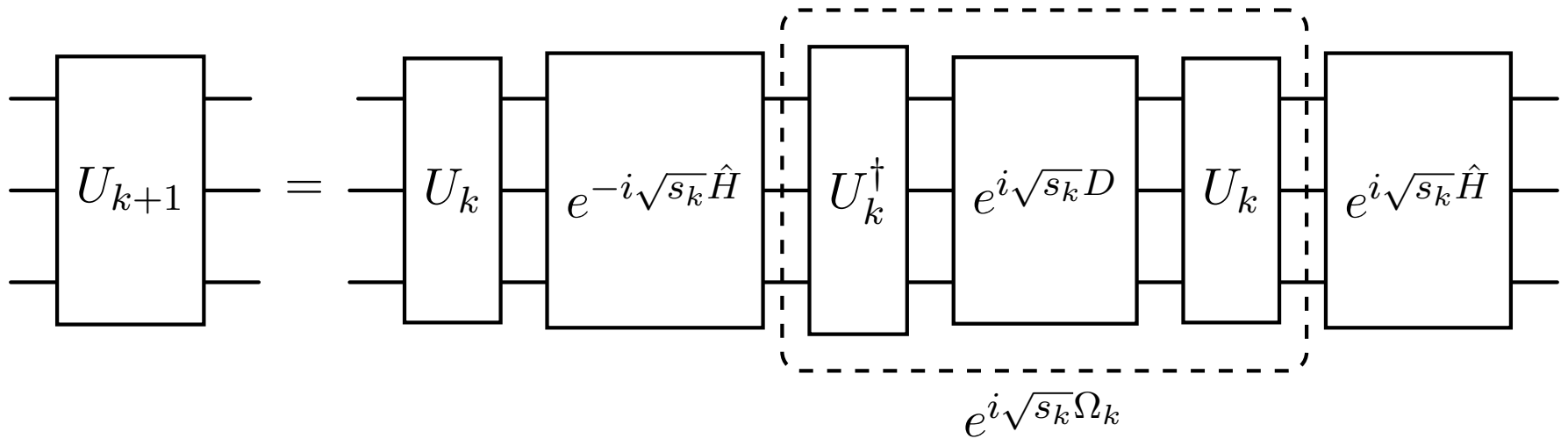
\begin{figure}[H]
\centering
\begin{quantikz}[row sep=0.3cm, column sep=0.35cm]
 & \gate[3]{U_{k+1}} & \midstick[3,brackets=none]{=}
 & \gate[3]{U_k} & \gate[3]{e^{-i\sqrt{s_k}\hat{H}}}
 & \gate[3]{U_k^\dagger}\gategroup[3,steps=3,style={dashed,rounded corners,inner xsep=2pt},label style={label position=below,anchor=north,yshift=-0.2cm}]{$e^{i\sqrt{s_k}\Omega_k}$}
 & \gate[3]{e^{i\sqrt{s_k}D}} & \gate[3]{U_k} & \gate[3]{e^{i\sqrt{s_k}\hat{H}}} & \\
 & & & & & & & & & \\
 & & & & & & & & &
\end{quantikz}
\caption{One recursion step of DB-MEX, Eq.~\eqref{eq:Uk}. The first $U_k$ prepares the current states ${\ket*{\phi_k^{(q)}}=U_k\ket{x_q}}$ from the inputs. The remaining gates apply the reduced group commutator~\eqref{eq:ket-update-GC}, in which the evolution under $\Omega_k$ (dashed box) is realized as $U_k\,e^{i\sqrt{s_k}D}\,U_k^\dagger$ by Eq.~\eqref{eq:factorization}. Because the generator $[\Omega_k,\hat H]$ depends on the current states, the circuit $U_k$ from the previous step enters $U_{k+1}$ three times.}
\label{fig:dbmex_circuit}
\end{figure}

Each step of the iteration~\eqref{eq:ket-update} applies the unitary $e^{s_k\comm{\Omega_k}{\hat H}}$, which contains the particular solution $\Omega_k$ for the initial ensemble operator. We now compile this unitary offline into a quantum circuit. We use the group-commutator approximation~\cite{gluza2026double}, which decomposes the exponential of a commutator into a sequence of evolutions under $\hat H$ and $\Omega_k$ with a local error of $\mathcal O(s_k^{3/2})$,
\begin{equation} \label{eq:gc}
    e^{s_k\comm{\Omega_k}{\hat H}}\approx e^{i\sqrt{s_k}\hat H}\, e^{i\sqrt{s_k}\Omega_k}\,e^{-i\sqrt{s_k}\hat H}\,e^{-i\sqrt{s_k}\Omega_k}.
\end{equation}
The guarantees of Sec.~\ref{sec:iteration} concern the exact step~\eqref{eq:ket-update}, and extending them to the compiled step, including the error of the group-commutator approximation, is left for future work. Higher-order group-commutator formulas can provide a better approximation (see Appendix~B.2 of Ref.~\cite{suzuki2025grover}), and other realizations of the step, for example, based on block encodings, are also possible. As in DB-QITE~\cite{gluza2026double}, when the group commutator~\eqref{eq:gc} is applied to a specific state $\ket*{\phi_k^{(q)}}$ in the ensemble, the rightmost term acts as a global phase on that state. Because $\ket*{\phi_k^{(q)}}$ is an eigenvector of $\Omega_k$ with eigenvalue $w_q$, we have
\begin{equation*}
    e^{-i\sqrt{s_k}\Omega_k}\ket*{\phi_k^{(q)}} =e^{-i\sqrt{s_k}\sum_{j=0}^p w_j\ket*{\phi_k^{(j)}}\bra*{\phi_k^{(j)}}} \ket*{\phi_k^{(q)}} =e^{-i\sqrt{s_k}w_q}\ket*{\phi_k^{(q)}}.
\end{equation*}
This phase depends on $q$, but each state enters $\Omega_{k+1}$ only through $\ketbra*{\phi^{(q)}_{k+1}}{\phi^{(q)}_{k+1}}$, where the phase cancels. Since each step depends on the states only through $\Omega_k$, discarding these phases leaves all later steps unchanged. By induction on $k$, the full and reduced recursions have the same $\Omega_k$ at every step, and their states differ only by a $q$-dependent phase. We therefore discard them and obtain the reduced recursion
\begin{equation} \label{eq:ket-update-GC}
    \ket*{\phi_{k+1}^{(q)}}=e^{i\sqrt{s_k}\hat H}\,e^{i\sqrt{s_k}\Omega_k}\, e^{-i\sqrt{s_k}\hat H}\,\ket*{\phi_k^{(q)}}.
\end{equation}
Let $U_k$ be the circuit after $k$ steps, which prepares every state from its input, ${\ket*{\phi_k^{(q)}}=U_k\ket{x_q}}$ for all $q$. Then $\Omega_k=U_kDU_k^\dagger$ with $D$ from Eq.~\eqref{eq:D}, and therefore
\begin{equation} \label{eq:factorization}
e^{i\sqrt{s_k}\Omega_k}=U_k\,e^{i\sqrt{s_k}D}\,U_k^\dagger .
\end{equation}
The update~\eqref{eq:ket-update-GC} applies the same unitary to every state. Starting from $U_0$, we define the circuits recursively by
\begin{equation}
    U_{k+1}:=e^{i\sqrt{s_k}\hat H}e^{i\sqrt{s_k}\Omega_k}e^{-i\sqrt{s_k}\hat H}\,U_k,
\end{equation}
which by~\eqref{eq:factorization} gives the recursive formula for unitary synthesis
\begin{equation} \label{eq:Uk}
    U_{k+1}=e^{i\sqrt{s_k}\hat H}\;U_k\,e^{i\sqrt{s_k}D}\,U_k^\dagger\;
    e^{-i\sqrt{s_k}\hat H}\;U_k,
\end{equation}
shown in Fig.~\ref{fig:dbmex_circuit}. Then $\ket*{\phi_{k+1}^{(q)}}=U_{k+1}\ket{x_q}$ for all $q$.

The circuit~\eqref{eq:Uk} has the same structure as the DB-QITE circuit~\cite{gluza2026double} and reduces exactly to it for $p=0$, $w_0=1$ and $\ket{x_0}=\ket{0\cdots0}$. The only difference is the weighted phase gate $e^{i\sqrt{s_k}D}$, which replaces the multi-controlled phase gate $e^{i\sqrt{s_k}\ketbra{0}{0}}$ of DB-QITE. App.~\ref{app:phase-gate} gives an implementation with Toffoli gates and single-qubit phase gates. For equal weights, it costs no more than the multi-controlled phase gate of DB-QITE. Targeted weights add at most $\lceil\log_2(p+1)\rceil$ Toffoli gates and one phase gate in total, and distinct weights at most $\lceil\log_2(p+1)\rceil$ Toffoli gates and one phase gate per additional state, as listed in Table~\ref{tab:gate-cost}. The real-time evolutions $e^{\pm i\sqrt{s_k}\hat H}$ are implemented using a Trotter--Suzuki decomposition~\cite{childs2021theory}. Since $U_K$ is unitary, the output states $U_K\ket{x_q}$ remain orthonormal despite the group-commutator and Trotter errors, which only change which states are prepared.

Since $U_k$ appears three times in $U_{k+1}$, the circuit $U_K$ contains $3^K$ copies of $U_0$, $3^K-1$ real-time evolutions and $(3^K-1)/2$ weighted phase gates, as in DB-QITE~\cite{gluza2026double}, so its depth grows as $\mathcal O(3^K)$ with the number of steps. This exponential growth is common to double-bracket algorithms compiled coherently and recursively with group commutators~\cite{gluza2026double,robbiati2026double,wright2026double}. In practice, they are therefore applied to tasks that require only a few steps, for example, refining a warm start already close to the target or providing a warm start for other algorithms.

Once $U_K$ is compiled, it is run only on the inputs whose states are needed. With distinct weights, preparing all eigenstates up to the $p$-th excited state requires running $U_K$ on all $p+1$ inputs, while with targeted weights, preparing only the $p$-th excited state requires the single run $U_K\ket{x_p}$. With equal weights, the outputs span the subspace of the $p+1$ lowest eigenstates but are in general not individual eigenstates, and Sec.~\ref{sec:postprocessing} describes how to resolve them.

\subsection{Resolving Eigenstates with Shared Weights}
\label{sec:postprocessing}
Let $U:=U_K$ and $\ket*{\phi^{(q)}}:=U\ket{x_q}$. As discussed in Sec.~\ref{sec:weights}, at convergence $[\Omega_K,\hat H]=0$, so $\hat H$ leaves every eigenspace of $\Omega_K$ invariant. An output state $\ket*{\phi^{(q)}}$ whose weight differs from all other weights is therefore an eigenstate of $\hat H$, while output states that share a weight span an invariant subspace but are in general not individual eigenstates. Post-processing is thus needed to resolve the individual eigenstates spanned by states that share a weight. As in multistate contracted VQE~\cite{parrish2019quantum}, where a shared circuit with equal weights is followed by diagonalizing $\hat H$ on its outputs, this reduces to diagonalizing a small matrix measured with the same circuit $U$.

For a set $\mathcal G$ of $d$ indices whose states share a weight, we construct the $d\times d$ matrix $H^{\mathcal G}$ by measuring its elements
\begin{equation} \label{eq:block-matrix}
    H^{\mathcal G}_{qq'}:=\bra*{\phi^{(q)}}\hat H\ket*{\phi^{(q')}}=\bra{x_q}U^\dagger\hat H\,U\ket{x_{q'}}
\end{equation}
for $q,q'\in\mathcal G$. For equal weights, $\mathcal G$ contains all $p+1$ states. The diagonal elements are expectation values. The off-diagonal elements are transition matrix elements, which can be estimated with a Hadamard test~\cite{cleve1998quantum} at the cost of an ancilla, or, more simply, from expectation values alone, with $d^2$ circuit settings in total (see Remark~\ref{rem:polarization} below).

Diagonalizing $H^{\mathcal G}$ classically gives its eigenpairs $(\tilde E_m,v^{(m)})$, with normalized eigenvectors $v^{(m)}\in\mathbb C^d$ of entries $v^{(m)}_q$, $q\in\mathcal G$. For each eigenpair, we define the input state $\ket*{v^{(m)}}:=\sum_{q\in\mathcal G}v^{(m)}_q\ket{x_q}$. To separate the individual eigenstates of $\hat H$ in the span of the output states $\ket*{\phi^{(q)}}$, $q\in\mathcal G$, we prepare the resolved states
\begin{equation} \label{eq:resolved-states}
    \ket*{\psi^{(m)}}:=\sum_{q\in\mathcal G}v^{(m)}_q\ket*{\phi^{(q)}}=\sum_{q\in\mathcal G}v^{(m)}_q\,U\ket{x_q}=U\ket*{v^{(m)}},  
\end{equation}
each obtained by preparing $\ket*{v^{(m)}}$ and applying $U$ once. At convergence, this span is invariant under $\hat H$, so
\begin{equation*}
    \hat H\ket*{\phi^{(q')}}=\sum_{q\in\mathcal G}H^{\mathcal G}_{qq'}\ket*{\phi^{(q)}}
\end{equation*}
for every $q'\in\mathcal G$, and therefore
\begin{equation*}
    \hat H\ket*{\psi^{(m)}} =\sum_{q,q'\in\mathcal G}H^{\mathcal G}_{qq'}v^{(m)}_{q'}\ket*{\phi^{(q)}} =\tilde E_m\sum_{q\in\mathcal G}v^{(m)}_q\ket*{\phi^{(q)}} =\tilde E_m\ket*{\psi^{(m)}} .
\end{equation*}
The $\ket*{\psi^{(m)}}$ are thus exact eigenstates of $\hat H$ with energies $\tilde E_m$.

At finite $K$, $\tilde E_m$ and $\ket*{\psi^{(m)}}$ are approximations whose accuracy can be assessed without knowing the spectrum. The energy of $\ket*{\psi^{(m)}}$ is $\tilde E_m$, and its energy variance
\begin{equation} \label{eq:variance}
    \bra*{\psi^{(m)}}\hat H^2\ket*{\psi^{(m)}}-\tilde E_m^2 =\big\|(\hat H-\tilde E_m)\ket*{\psi^{(m)}}\big\|^2 ,
\end{equation}
which is measured with the same circuit, vanishes if and only if $\ket*{\psi^{(m)}}$ is an eigenstate of $\hat H$. Expanding $\ket*{\psi^{(m)}}=\sum_j c_j\ket{E_j}$ in orthonormal eigenstates of $\hat H$ gives
\begin{equation*}
    \big\|(\hat H-\tilde E_m)\ket*{\psi^{(m)}}\big\|^2=\sum_j|c_j|^2(E_j-\tilde E_m)^2 \ge\min_j(E_j-\tilde E_m)^2 .
\end{equation*}
The square root of the variance therefore bounds the distance from $\tilde E_m$ to the nearest eigenvalue of $\hat H$, which provides a convergence check that does not require knowing the eigenstates. For outputs with distinct weights, which become eigenstates at convergence, the variance can be evaluated directly on them to provide the same check.

\begin{remark}\label{rem:polarization}
Since all output states are prepared by the same circuit $U$, the off-diagonal elements of $H^{\mathcal G}$ can be obtained from expectation values alone, as in subspace-search VQE~\cite[Sec.~IV]{nakanishi2019subspace}. With the auxiliary states
\begin{equation*}
    \ket*{\chi^{\mathrm R}_{qq'}}:=\tfrac{1}{\sqrt2}U\big(\ket{x_q}+\ket{x_{q'}}\big),
    \qquad
    \ket*{\chi^{\mathrm I}_{qq'}}:=\tfrac{1}{\sqrt2}U\big(\ket{x_q}-i\ket{x_{q'}}\big),
\end{equation*}
and their energies $H^{\mathrm R}_{qq'}:=\bra*{\chi^{\mathrm R}_{qq'}}\hat H\ket*{\chi^{\mathrm R}_{qq'}}$ and $H^{\mathrm I}_{qq'}:=\bra*{\chi^{\mathrm I}_{qq'}}\hat H\ket*{\chi^{\mathrm I}_{qq'}}$,
\begin{equation} \label{eq:polarization}
    \mathrm{Re}\big(H^{\mathcal G}_{qq'}\big)=H^{\mathrm R}_{qq'}-\tfrac12H^{\mathcal G}_{qq}-\tfrac12H^{\mathcal G}_{q'q'},
    \qquad
    \mathrm{Im}\big(H^{\mathcal G}_{qq'}\big)=H^{\mathrm I}_{qq'}-\tfrac12H^{\mathcal G}_{qq} -\tfrac12H^{\mathcal G}_{q'q'}.
\end{equation}
\end{remark}

Algorithm~\ref{alg:dbmex} summarizes the full procedure, from circuit synthesis in Sec.~\ref{sec:circuit} to post-processing.

\begin{algorithm}[H]
\caption{DB-MEX}
\label{alg:dbmex}
\begin{algorithmic}[1]
\Require  Hamiltonian $\hat H$, weights $w_0\ge\dots\ge w_p>0$, distinct computational basis states $\ket{x_0},\dots,\ket{x_p}$, initial circuit $U_0$, number of steps $K$, step sizes $s_0,\dots,s_{K-1}$
\Ensure Circuit $U_K$, approximate eigenstates of $\hat H$, and their energies
\State $D\gets\sum_{q=0}^pw_q\ket{x_q}\bra{x_q}$ \Comment{Eq.~\eqref{eq:D}}
\For{$k=0,\dots,K-1$}
  \State $U_{k+1}\gets e^{i\sqrt{s_k}\hat H}\,U_k\,e^{i\sqrt{s_k}D}\, U_k^\dagger\,e^{-i\sqrt{s_k}\hat H}\,U_k$ \Comment{Eq.~\eqref{eq:Uk}} 
\EndFor
\State For each state with its own weight, prepare $U_K\ket{x_q}$ and measure its energy
\ForAll{sets $\mathcal G$ of indices whose states share a weight}
  \State Measure $H^{\mathcal G}$ with     Eqs.~\eqref{eq:block-matrix} and~\eqref{eq:polarization}
  \State Diagonalize $H^{\mathcal G}$ to obtain the pairs
    $(\tilde E_m,v^{(m)})$
  \State Prepare $\ket*{\psi^{(m)}}=U_K\ket*{v^{(m)}}$
    \Comment{Eq.~\eqref{eq:resolved-states}}
\EndFor
\end{algorithmic}
\end{algorithm}

\section{Conclusion}
\label{sec:conclusion}
We introduce DB-MEX, a double-bracket quantum algorithm that targets several low-lying eigenstates of a Hamiltonian with one recursively compiled circuit. The choice of weights selects whether it targets the $p+1$ lowest eigenstates, only the $p$-th excited state, or only their span, which a post-processing step resolves. We prove that the double-bracket flow of the ensemble operator, whose rank and weights are design choices, converges almost globally to its target, with explicit local rates given by products of weight and energy gaps. We also show that its discretization decreases the weighted energy at every step and, for small step sizes, converges to the same target.

The structure of the compiled circuit differs from that of DB-QITE only in the weighted phase gate, for which we give an explicit implementation. Since all states are produced by the same circuit, they are orthonormal by construction, which also simplifies estimating transition amplitudes between them, as in Remark~\ref{rem:polarization}. Such amplitudes are useful, for example, for computing absorption spectra~\cite{parrish2019quantum}.

As in DB-QITE and other recursively compiled double-bracket algorithms, the circuit depth grows as $\mathcal O(3^K)$ with the number of steps, so DB-MEX is best suited to tasks that need only a few steps, for example, to refine a warm start that is already close to the target. A probabilistic variant of DB-QITE with exponentially shorter circuits has been proposed~\cite{kolotouros2026quantum}. Its per-step success probability can remain close to one, but the total success probability decays exponentially with the number of steps. Whether a similar variant exists for DB-MEX is left for future work.

The flow underlying DB-MEX generalizes the double-bracket formulation of ITE used in DB-QITE~\cite{gluza2026double}. It reduces to ITE for $p=0$, but for $p>0$ it is in general no longer an ITE of the individual states. More broadly, the analysis of DB-MEX builds on established results about gradient flows on matrix manifolds from control and dynamical systems theory, which provide convergence guarantees, limit sets and rates, and we expect these results to be useful for designing further quantum algorithms whose underlying flows have provable convergence.

\bibliographystyle{quantum}
\bibliography{references}

\onecolumn
\appendix
\newpage

\section{Convergence of the Flow}
\label{app:morse-bott}
This appendix proves the convergence of the flow~\eqref{eq:dbf} and its local rate, Theorems~\ref{thm:flow} and~\ref{thm:local-rate}, for arbitrary multiplicities in the spectra of $\Omega$ and $\hat H$. The proof builds on established results on gradient flows on matrix manifolds~\cite{helmke2012optimization} and is self-contained.

The appendix is organized as follows. Sec.~\ref{sec:AppendixPreliminaries} recalls Morse--Bott functions and the properties of their gradient flows that we use. Sec.~\ref{sec:AppendixTheorem1} shows that $V$ is a Morse--Bott function on $M(\bar\Omega)$ with a unique minimum manifold and proves Theorem~\ref{thm:flow}. Sec.~\ref{sec:app-rates} computes the local rate and proves Theorem~\ref{thm:local-rate}.

\subsection{Preliminaries}
\label{sec:AppendixPreliminaries}

In this section, we consider gradient flows for Morse--Bott functions. In particular, we define critical, minimum, stable/unstable manifolds and state some lemmas to be used in the next section to prove the convergence theorem for such flows.

Let $M$ be a smooth manifold of dimension $n$ and $V \colon M \to \R$
be a smooth ($C^\infty$) function. We write
\[
  \Crit(V) = \{ p \in M : dV_p = 0 \}
\]
for the set of critical points of $V$. At a critical point $p$, the \emph{Hessian} $\Hess_p V \colon T_pM \times T_pM \to \R$ is the well-defined symmetric bilinear form given in local coordinates by $\Hess_p V(u,v) = \sum_{i,j} \frac{\partial^2 V}{\partial x_i \partial x_j}(p)\, u_i v_j$.

\begin{definition}[Nondegenerate critical submanifold]
\label{def:nondeg-crit-mfd}
A connected submanifold $C \subseteq M$ is a \emph{critical submanifold} of $V$ if $C \subseteq \Crit(V)$. It is called \emph{nondegenerate} (in the sense of Bott) if, for every $p \in C$,
\[
  T_pC = \ker \Hess_p V ,
  \qquad
  \ker \Hess_p V = \{ u \in T_pM : \Hess_p V(u,v) = 0 \ \ \forall v \in T_pM \}.
\]
\end{definition}

Equivalently, $\Hess_p V$ can be thought of as a nondegenerate quadratic form on the normal space $N_pC = T_pM / T_pC$.

\begin{definition}[Morse--Bott function]
\label{def:morse-bott}
The function $V$ is a \emph{Morse--Bott function} if $\Crit(V)$ is a disjoint union of nondegenerate critical submanifolds.
\end{definition}

\begin{definition}[Index and normal bundles]
\label{def:index}
Let $C$ be a nondegenerate critical submanifold and $NC \to C$ its normal bundle. The nondegenerate form induced by $\Hess V$ on $NC$ yields a splitting
\[
  NC = N^-C \oplus N^+C
\]
into subbundles on which $\Hess V$ is negative, respectively positive, definite. The \emph{index} of $C$ is $\lambda_C = \rank N^-C$, and the \emph{coindex} is $\rank N^+C = \codim C - \lambda_C$. The bundle $N^-C$ is called the \emph{negative normal bundle} of $C$.
\end{definition}

\begin{definition}[Minimum manifold]
A nondegenerate critical submanifold $C$ is a \emph{minimum manifold} of $V$ if $NC = N^+C$, i.e.\ $\lambda_C = 0$. 
\end{definition}

The following lemmas are used to characterize Morse--Bott functions near a nondegenerate critical manifold.

\begin{theorem}[Morse--Bott lemma\cite{BanyagaHurtubise2004}, local form]
\label{thm:mb-lemma-local}
Let $C \subseteq M$ be a nondegenerate critical submanifold of $V$ with $\dim C = c$ and index $\lambda$. Then for every $p \in C$ there exist a neighbourhood $U$ of $p$ and local coordinates
\[
  (x, y, z) = (x_1,\dots,x_c,\, y_1,\dots,y_\lambda,\, z_1,\dots,z_{n-c-\lambda})
  \colon U \to \R^n
\]
centred at $p$, with $C \cap U = \{ y = 0,\ z = 0 \}$, such that
\[
  V(x,y,z) = V(C) - \sum_{i=1}^{\lambda} y_i^2 + \sum_{j=1}^{n-c-\lambda} z_j^2
  \qquad \text{on } U .
\]
\end{theorem}

\begin{theorem}[Morse--Bott lemma, global form]
\label{thm:mb-lemma-global}
Let $C$ be a compact nondegenerate critical submanifold of $V$, and fix a fiber metric on $NC = N^-C \oplus N^+C$. Then there exist a neighbourhood $U$ of the zero section in $NC$ and a diffeomorphism $\phi$ from $U$ onto a neighbourhood of $C$ in $M$, restricting to the identity on $C$, such that
\[
  (V \circ \phi)(v_-, v_+) = V(C) - \lvert v_- \rvert^2 + \lvert v_+ \rvert^2,
  \qquad (v_-, v_+) \in U \cap (N^-C \oplus N^+C).
\]
\end{theorem}

Let us fix a Riemannian metric $g$ on $M$ and let $\nabla V$ denote the gradient of $V$ with respect to $g$. We consider the gradient descent flow
\begin{equation} \label{eq:gradient-flow}
  \dot x = -\nabla V(x), \qquad x(0) = x_0 \in M,
\end{equation}
and denote its flow by $\varphi_t$. Along solutions, $\frac{d}{dt} V(x(t)) = -\lVert \nabla V(x(t)) \rVert_g^2 \le 0$, with equality only at critical points.

\begin{definition}[Stable manifold and basin of attraction]
For a critical submanifold $C$ of $V$, the \emph{stable set} of $C$ is
\[
  W^s(C) = \bigl\{ x_0 \in M : \varphi_t(x_0) \text{ is defined for all } t \ge 0  \text{ and } \lim_{t \to \infty} \operatorname{dist}(\varphi_t(x_0), C) = 0 \bigr\}.
\]
\end{definition}

The next lemma is a consequence of the Morse--Bott lemma.

\begin{lemma}[Łojasiewicz inequality near a critical manifold]
\label{lem:lojasiewicz}
Let $C$ be a compact nondegenerate critical submanifold of $V$. Then there exist a neighbourhood $U$ of $C$ and a constant $\kappa > 0$ such that
\[
  \lvert V(x) - V(C) \rvert^{1/2} \le \kappa \, \lVert \nabla V(x) \rVert_g
  \qquad \text{for all } x \in U .
\]
\end{lemma}

\begin{proof}
In the coordinates of \cref{thm:mb-lemma-global}, $\lvert V - V(C) \rvert \le \lvert v_- \rvert^2 + \lvert v_+ \rvert^2$ while $\lVert dV \rVert \ge 2\,(\lvert v_- \rvert^2 + \lvert v_+ \rvert^2)^{1/2}$ up to constants depending on $\phi$ and $g$, which are uniform on a compact neighbourhood of $C$.
\end{proof}

\begin{proposition}[Stable manifold theorem]
\label{prop:stable-manifolds}
Let $C$ be a compact nondegenerate critical submanifold of $V$. Then $C$ is a normally hyperbolic invariant manifold of \eqref{eq:gradient-flow}, and $W^s(C)$ is a countable union of embedded submanifolds of dimension $\dim C+\rank N^+C$. In particular, if $\lambda_C>0$, then $W^s(C)$ has Lebesgue measure zero in $M$.
\end{proposition}

\subsection{Proof of Theorem \ref{thm:flow}}
\label{sec:AppendixTheorem1}
To prove Theorem~\ref{thm:flow}, we first consider a generalization of \cite[Section~2, Theorem~1.5]{helmke2012optimization} to the case of Hermitian matrices with possibly repeated eigenvalues.
\begin{theorem}[Almost global convergence to the minimum manifold]
\label{thm:almost-global}
Let $M$ be compact, let $V \colon M \to \R$ be a Morse--Bott function with critical submanifolds $C_1, \dots, C_m$, and suppose that exactly one of them, $C_* := C_1$, is a minimum manifold. Then, for the gradient flow
\eqref{eq:gradient-flow}:
\begin{enumerate}
  \item every solution $x(t)$ converges, as $t \to \infty$, to a single point $x_\infty \in \Crit(V)$, and the convergence is exponential;
  \item almost every solution $x(t)$ converges to $C_*$.
\end{enumerate}
\end{theorem}

\begin{proof}
\emph{(1)} Since $M$ is compact the flow is complete, and $V$ has compact level sets. Hence, \cite[Appendix C, Proposition~12.3]{helmke2012optimization} implies convergence to a single equilibrium. The function $V$ is a Lyapunov function. By LaSalle's invariance principle, the $\omega$-limit set $\omega(x_0)$ is a nonempty, compact, connected subset of $\Crit(V)$; since the $C_j$ are disjoint and compact, $\omega(x_0) \subseteq C_j$ for a single $j$. Let $p \in \omega(x_0)$. Then $V(x(t)) \searrow V(C_j)$. If $V(x(t_0)) = V(C_j)$ for some $t_0$, then $x(t_0)$ is an equilibrium and we are done. Otherwise, set $h(t) = \bigl(V(x(t))-V(C_j) \bigr)^{1/2} > 0$. While $x(t)$ lies in the neighbourhood $U$ of \cref{lem:lojasiewicz},
\[
  \dot h = -\frac{\lVert \nabla V(x) \rVert^2}{2h}
  \le -\frac{1}{2\kappa}\, \lVert \nabla V(x) \rVert
  = -\frac{1}{2\kappa}\, \lVert \dot x \rVert .
\]
Combining $\dot h \le -\frac{1}{2\kappa} \lVert \nabla V \rVert$ with \cref{lem:lojasiewicz} gives $\dot h \le -\frac{1}{2\kappa^2} h$, i.e.\ exponential convergence.

\emph{(2)} By (1), every solution converges to exactly one critical submanifold. Hence, $M = W^s(C_*) \,\sqcup\, \bigsqcup_{j=2}^{m} W^s(C_j)$, and each $C_j$ with $j \ge 2$ satisfies $N^-C_j \neq 0$ by uniqueness of the minimum manifold. Hence, by \cref{prop:stable-manifolds}, $\bigsqcup_{j=2}^{m} W^s(C_j)$ has measure zero and therefore $W^s(C_*)$ has full measure.
\end{proof}
We now consider the particular case where the gradient flow considered in the above theorem is a double bracket flow with repeated eigenvalues. Let $\operatorname{Herm}(L)$ denote the real vector space of complex Hermitian $L \times L$ matrices with inner product $\langle A, B \rangle = \operatorname{tr}(A B)$, and let $\mathfrak{u}(L)$ denote the skew-Hermitian matrices, the Lie algebra of the unitary group $\mathrm{U}(L)$, with the real inner product $\langle \Omega_1, \Omega_2 \rangle = -\operatorname{tr}(\Omega_1 \Omega_2)$. Fix $\bar\Omega \in \operatorname{Herm}(L)$ with (real) eigenvalues
\[
  w_0 \ge w_1 \ge \dots \ge w_{L-1}
\]
(repetitions allowed), and let
\[
  M(\bar\Omega) = \{ U \bar\Omega U^\dagger : U \in \mathrm{U}(L) \}
\]
be the isospectral manifold. As an orbit of the compact connected Lie group $\mathrm{U}(L)$, it is a compact connected embedded submanifold of $\operatorname{Herm}(L)$. For a given $\hat H \in \operatorname{Herm}(L)$ consider the double bracket flow
\begin{equation}
  \label{eq:double-bracket}
  \dot \Omega = [[\Omega, \hat H],\Omega], \qquad \Omega(0) = \Omega_0 \in M(\bar\Omega),
\end{equation}
and the function $V \colon M(\bar\Omega) \to \R$, $V(\Omega) = \operatorname{tr}(\hat H\Omega)$. Since $\lVert \Omega \rVert_{HS}$ is constant on $M(\bar\Omega)$, $V$ differs from $-\tfrac12 \lVert \hat H - \Omega \rVert_{HS}^2$ by a constant. 

We now show that, for the limit behavior of the above flow, only the ordering and multiplicities of the eigenvalues matter:
\begin{enumerate}
  \item If $\hat H = U \tilde H U^\dagger$ with $U \in \mathrm{U}(L)$, then $\Omega(t)$ solves \eqref{eq:double-bracket} if and only if $U^\dagger \Omega(t) U$ solves it with $\hat H$ replaced by $\tilde H$. Hence we may assume $\hat H$ is a real diagonal matrix.
  \item For $\alpha > 0$ and $\beta \in \R$ we have $[[\Omega,\alpha \hat H + \beta I],\Omega] = \alpha [[\Omega,\hat H],\Omega]$, so replacing $\hat H$ by $\alpha \hat H + \beta I$ only rescales time.
\end{enumerate}
Consequently, the values of the eigenvalues of $\hat H$ are irrelevant for the limit behavior; only their \emph{ordering} and \emph{multiplicities} matter. In particular, a repeated eigenvalue $0$ plays no special role. Accordingly, we assume from now on
\begin{equation} \label{eq:N-normal-form}
  \hat H = \operatorname{diag}\bigl( \nu_1 I_{n_1}, \nu_2 I_{n_2}, \dots, \nu_r I_{n_r} \bigr),  \qquad \nu_1 < \nu_2 < \dots < \nu_r ,
\end{equation}
with multiplicities $n_1 + \dots + n_r = L$. Put $m_0 = 0$, $m_k = n_1 + \dots + n_k$, let $\mathcal{I}_k = \{ m_{k-1}+1, \dots, m_k \}$ be the $k$-th block of indices, and write $\nu(p)=\nu_k=E_{p-1}$ for $p \in \mathcal{I}_k$. Let
\[
  K = \mathrm{U}(n_1) \times \dots \times \mathrm{U}(n_r) \subseteq \mathrm{U}(L)
\]
(embedded block-diagonally); it is compact and connected, and its elements commute with $\hat H$.

\begin{definition}[Block spectra and the sorted critical manifold]
\label{def:block-spectra}
Let $\mathcal{D}$ be the finite set of real diagonal matrices $D = \operatorname{diag}(d_1, \dots, d_L)$ whose diagonal is a permutation of $(w_0, \dots, w_{L-1})$. Call $D, D' \in \mathcal{D}$ \emph{equivalent} if they differ by a permutation of entries within each block $\mathcal{I}_k$, and for $D \in \mathcal{D}$ put
\[
  C_D = \{ \Theta D \Theta^\dagger : \Theta \in K \}.
\]

Let $D_* = \operatorname{diag}(w_0, \dots, w_{L-1})$ and
\[
  C_* := C_{D_*}
  = \Bigl\{ \operatorname{diag}(H_1, \dots, H_r) :
       H_k \in \operatorname{Herm}(n_k),\
       \operatorname{spec}(H_k) = (w_{m_{k-1}}, \dots, w_{m_k-1})
    \Bigr\}.
\]
\end{definition}

Thus $C_*$ consists of the block-diagonal Hermitian matrices in which the $k$-th block carries the $k$-th group of eigenvalues of $\bar\Omega$ in decreasing order: the largest $n_1$ eigenvalues in the block belonging to the smallest eigenvalue $\nu_1$ of $\hat H$, the next $n_2$ in the second block, and so on.

To prove Theorem~\ref{thm:flow}, we need the following lemmas:

\begin{lemma}[Gradient structure]
\label{lem:db-gradient}
Equip $M(\bar\Omega)$ with the normal metric: for $\Omega\in M(\bar\Omega)$ and $\Omega_1, \Omega_2 \in \mathfrak{u}(L)$,
\[
  \bigl\langle [\Omega_1, \Omega], [\Omega_2, \Omega] \bigr\rangle_\Omega
  = -\operatorname{tr}\bigl( \Omega_1^\perp \Omega_2^\perp \bigr),
\]
where $\star^\perp$ is the orthogonal projection of $\star$ onto the orthogonal complement in $\mathfrak{u}(L)$ of the stabiliser $\mathfrak{u}_\Omega := \{ H \in \mathfrak{u}(L) : [\Omega, H] = 0 \}$. Then $\nabla V(\Omega) = [\Omega,[\Omega,\hat H]]$, and along solutions of
\eqref{eq:double-bracket}
\[
  \frac{d}{dt} V(\Omega(t)) = -\lVert [\Omega(t), \hat H] \rVert_{HS}^2 .
\]
\end{lemma}

\begin{proof}
Tangent vectors at $\Omega$ are of the form $[H,\Omega]$ with $H \in \mathfrak{u}(L)$, and $DV(\Omega)[H, \Omega] = \operatorname{tr}(\hat H[H,\Omega]) = \operatorname{tr}([\Omega,\hat H]H)$, a real number. Put $X = [\hat H, \Omega] \in \mathfrak{u}(L)$. If $H \in \mathfrak{u}_\Omega$ then $\operatorname{tr}(X H) = \operatorname{tr}(\hat H[\Omega,H]) = 0$, so $X \perp \mathfrak{u}_\Omega$, i.e.\ $X = X^\perp$, and now take an arbitrary $R \in \mathfrak{u}(L)$
\[
  \bigl\langle [X, \Omega], [R,\Omega] \bigr\rangle_\Omega
  = -\operatorname{tr}(X R^\perp)
  = -\operatorname{tr}(X R)
  = \operatorname{tr}([\Omega,\hat H]R)
  = DV(\Omega)[R,\Omega].
\]
Hence $\nabla V(\Omega) = [X, \Omega] = [[\hat H,\Omega],\Omega] = [\Omega,[\Omega,\hat H]]$. Finally, with $A := [\Omega,\hat H] \in \mathfrak{u}(L)$, $\frac{d}{dt} V = -\operatorname{tr}(\hat H [\Omega,[\Omega,\hat H]]) = -\operatorname{tr}([\hat H,\Omega][\Omega,\hat H]) = \operatorname{tr}(A^2) = -\operatorname{tr}(A^\dagger A) = -\lVert [\Omega,\hat H] \rVert_{HS}^2$.
\end{proof}

\begin{lemma}[Critical points]
\label{lem:db-critical}
$\Omega \in M(\bar\Omega)$ is an equilibrium of \eqref{eq:double-bracket} if and only if $[\Omega, \hat H] = 0$, if and only if $\Omega$ is block-diagonal with respect to the blocks $\mathcal{I}_1, \dots, \mathcal{I}_r$. Moreover,
\[
  \Crit(V) = \bigsqcup_{[D]} C_D ,
\]
the disjoint union over equivalence classes $[D]$ of elements of $\mathcal{D}$, and each $C_D$ is a compact, connected, embedded submanifold of $M(\bar\Omega)$.
\end{lemma}

\begin{proof}
If $[\Omega,[\Omega,\hat H]] = 0$ then $\lVert [\Omega,\hat H] \rVert_{HS}^2 = \operatorname{tr}(\hat H[\Omega,[\Omega,\hat H]]) = 0$ by \cref{lem:db-gradient}; the converse is trivial. A Hermitian matrix commutes with $\hat H$ in \eqref{eq:N-normal-form} if and only if it is block-diagonal. Diagonalising each Hermitian block by a unitary matrix gives $\Omega = \Theta D \Theta^\dagger$ with $D \in \mathcal{D}$ and $\Theta \in K$, so $\Omega \in C_D$. The multiset of eigenvalues of each block of $\Omega$ is an invariant, so $C_D = C_{D'}$ if $D$ and $D'$ are equivalent, and $C_D \cap C_{D'} = \emptyset$ otherwise. Each $C_D$ is an orbit of the compact connected Lie group $K$, hence a compact connected embedded submanifold.
\end{proof}

\begin{lemma}[Hessian at a diagonal critical point]
\label{lem:db-hessian}
Let $D = \operatorname{diag}(d_1, \dots, d_L) \in \mathcal{D}$. Then
\[
  T_D M(\bar\Omega) = \{ \xi \in \operatorname{Herm}(L) : \xi_{pq} = 0 \text{ whenever } d_p = d_q \},
\]
so the real and imaginary parts of the complex entries $\xi_{pq}$, $p < q$, $d_p \neq d_q$, are real linear coordinates on $T_D M(\bar\Omega)$, and
\begin{equation}
  \label{eq:db-hessian}
  \Hess_DV(\xi, \xi)
  = 2 \sum_{\substack{p < q \\ d_p \neq d_q}}
      \frac{\nu(p) - \nu(q)}{d_q - d_p}\, \lvert \xi_{pq} \rvert^2 .
\end{equation}
\end{lemma}
\begin{proof}
For $H \in \mathfrak{u}(L)$ we have $[H, D]_{pq} = H_{pq}(d_q - d_p)$, and $H_{pq} \in \mathbb{C}$ ($p < q$) can be chosen freely; this gives the description of $T_D M(\bar\Omega) = \{ [H, D] : H \in \mathfrak{u}(L) \}$. Since $D$ is a critical point, the second derivative of $V$ along any curve through $D$ with velocity $\xi$ equals $\Hess_D V(\xi,\xi)$. Take $\gamma(\theta) = e^{\theta H} D e^{-\theta H}$, so that $\gamma'(0) = [H, D] = \xi$ and
\[
  \Hess_D V(\xi,\xi)
  = \operatorname{tr}\bigl(\hat H [H, [H, D]]\bigr)
  = -\operatorname{tr}\bigl([H, \hat H][H, D]\bigr).
\]
With $[H, \hat H]_{pq} = H_{pq}(\nu(q) - \nu(p))$ and $H_{qp} = -\overline{H_{pq}}$, so that $[H, D]_{qp} = \overline{H_{pq}}\,(d_q - d_p)$, this equals $-\sum_{p,q} \lvert H_{pq} \rvert^2 (\nu(q) - \nu(p))(d_q - d_p) = -2 \sum_{p<q} \lvert H_{pq} \rvert^2 (\nu(p) - \nu(q))(d_p - d_q)$. Substituting $\lvert H_{pq} \rvert = \lvert \xi_{pq} \rvert / \lvert d_q - d_p \rvert$ for $d_p \neq d_q$ (the terms with $d_p = d_q$ vanish) gives
\eqref{eq:db-hessian}.
\end{proof}

\begin{proposition}[$V$ is Morse--Bott with a unique minimum manifold]
\label{prop:db-morse-bott}
The function $V$ is a Morse--Bott function on $M(\bar\Omega)$ whose critical submanifolds are the $C_D$. For $D \in \mathcal{D}$,
\[
  \rank N^+ C_D
  = 2\, \#\{ (p,q) : p < q,\ \nu(p) < \nu(q),\ d_q < d_p \}.
\]
Similarly, $\lambda_{C_D} = \rank N^- C_D$ is twice the number of pairs $p < q$ with $\nu(p) < \nu(q)$ and $d_p < d_q$. In particular, all indices are even, and $C_D$ is a minimum manifold if and only if $C_D = C_*$.
\end{proposition}

\begin{proof}
\emph{Nondegeneracy at diagonal points.} The tangent space of the orbit $C_D$ at $D$ is $\{ [H, D] : H \in \mathfrak{u}(L) \text{ block-diagonal} \}$, i.e.\ the span of the coordinates $\xi_{pq}$ with $p, q$ in the same block and $d_p \neq d_q$. By \eqref{eq:db-hessian}, the coefficient of $\lvert \xi_{pq} \rvert^2$ vanishes exactly when $\nu(p) = \nu(q)$, i.e.\ when $p$ and $q$ lie in the same block. Since the Hessian is diagonal in these coordinates, $\ker \Hess_D V = T_D C_D$.

\emph{Nondegeneracy at all points.} Any $\Omega \in C_D$ is $\Omega = \Theta D \Theta^\dagger$ with $\Theta \in K$. The diffeomorphism $\psi(X) = \Theta X \Theta^\dagger$ of $M(\bar\Omega)$ maps $C_D$ onto itself and satisfies $V \circ \psi = V$, because $\Theta^\dagger \hat H \Theta = \hat H$. Hence $\psi$ carries $\ker \Hess_D V = T_D C_D$ onto $\ker \Hess_\Omega V = T_\Omega C_D$.

\emph{Index.} In the coordinates of \cref{lem:db-hessian} the Hessian on the normal space is diagonal. The coefficient of $\lvert \xi_{pq} \rvert^2$ ($p < q$, $\nu(p) < \nu(q)$) is positive precisely when $d_p > d_q$ and negative precisely when $d_p < d_q$. Each complex coordinate $\xi_{pq}$ contributes two real dimensions, which gives the formulas for $\rank N^\pm C_D$.

\emph{Uniqueness of the minimum manifold.} $\rank N^- C_D = 0$ if and only if every entry of $D$ in block $\mathcal{I}_i$ is larger than or equal to every entry in block $\mathcal{I}_j$ whenever $i < j$. Then block $\mathcal{I}_1$ carries the $n_1$ largest eigenvalues of $\bar\Omega$ (as a multiset), block $\mathcal{I}_2$ the next $n_2$, and so on; thus $D$ is equivalent to $D_*$ and $C_D = C_*$.
\end{proof}

\begin{proof}[Proof of \cref{thm:flow}]
$M(\bar\Omega)$ is compact, and by \cref{lem:db-gradient} the flow \eqref{eq:double-bracket} is the gradient descent flow of $V$ for the normal metric. In particular, it preserves $M(\bar\Omega)$ and is complete. By \cref{lem:db-critical,prop:db-morse-bott}, $V$ is Morse--Bott with critical submanifolds $C_D$, of which exactly one, $C_*$, is a minimum manifold. \Cref{thm:almost-global} now gives exponential convergence of every solution to a single equilibrium $\Omega_\infty$, which is block-diagonal by \cref{lem:db-critical}, and convergence to $C_*$ for all $\Omega(0)$ in an open, dense, full-measure set whose complement is $\bigcup_{C_D \neq C_*} W^s(C_D)$. By \cref{prop:stable-manifolds} each $W^s(C_D)$ in this union has real codimension $\rank N^-C_D \ge 2$, since $\rank N^-C_D \neq 0$ for $C_D \neq C_*$ and it is even.
\end{proof}

\subsection{Local Rates}
\label{sec:app-rates}
The constant $\kappa$ of \cref{lem:lojasiewicz} is not explicit. Here we compute the rate near $C_*$. Notice that the coefficients in \eqref{eq:db-hessian} are taken in the Euclidean coordinates $\xi$ and they give the index, but not the rate, which is measured in the normal metric of \cref{lem:db-gradient}.

We keep the setting of \cref{sec:AppendixTheorem1}. For $1\le m<n\le L$ put
\begin{equation}
  \label{eq:eta}
  \eta_{mn} := \bigl(w_{m-1}-w_{n-1}\bigr)\bigl(E_{n-1}-E_{m-1}\bigr) \ge 0 ,
\end{equation}
and assume that $V$ is not constant, so that
\begin{equation*}
  \mu_* = \min\{ \eta_{mn} : m<n,\ \eta_{mn}>0 \}
\end{equation*}
is defined. We call a pair $m<n$ \emph{normal} if $\eta_{mn} > 0$, i.e.\ if $E_{m-1} \neq E_{n-1}$ and $w_{m-1} \neq w_{n-1}$. Recall that $V_*=V(C_*)$ is the minimum of $V$.

\begin{lemma}[Hessian at $C_*$ in the normal metric]
\label{lem:normal-hessian}
For $R \in \mathfrak{u}(L)$ and $\Omega = e^{R} D_* e^{-R}$,
\[
  V(\Omega) = V(D_*) + \sum_{m<n} \eta_{mn} \lvert R_{mn} \rvert^2 + O(\|R\|^3),
  \qquad
  \|[\Omega, \hat H]\|_{HS}^2 = 2 \sum_{m<n} \eta_{mn}^2 \lvert R_{mn} \rvert^2 + O(\|R\|^3).
\]
Hence the linearization of \eqref{eq:double-bracket} at $D_*$ is $\dot R_{mn} = -\eta_{mn}\,R_{mn}$.
\end{lemma}

\begin{proof}
Since $D_*$ is critical, the first formula is the second order term $\tfrac12 \operatorname{tr}(\hat H[R,[R,D_*]])$ computed in the proof of \cref{lem:db-hessian}. Since $[D_*,\hat H]=0$, $[\Omega,\hat H] = [[R,D_*],\hat H] + O(\|R\|^2)$ with $(m,n)$-entries $R_{mn}(w_{n-1}-w_{m-1})(E_{n-1}-E_{m-1})$. In the normal metric a single pair has squared speed $2\lvert R_{mn}\rvert^2$ and second derivative of $V$ equal to $2\eta_{mn}\lvert R_{mn}\rvert^2$, so the Hessian operator is $\eta_{mn}$ on that pair.
\end{proof}

\begin{theorem}[Local rate of convergence to $C_*$]
\label{thm:local-rate-general}
For every $0 < c < 2\mu_*$ there exist an invariant neighborhood $\mathcal U$ of $C_*$ and a constant $\gamma>0$ such that
\begin{enumerate}
\item $\|[\Omega,\hat H]\|_{HS}^2 \ge c\,( V(\Omega) - V_* )$ for all $\Omega \in \mathcal U$;
\item if $\Omega(t_0) \in \mathcal U$, then $V(\Omega(t)) - V_* \le e^{-c(t-t_0)} ( V(\Omega(t_0)) - V_* )$ for all $t \ge t_0$;
\item if $\Omega(t_0) \in \mathcal U$, then $\Omega(t)$ converges to a point $\Omega_\infty\in C_*$, and 
        \[
  \|\Omega(t)-\Omega_\infty\|_{HS} \le \gamma\, e^{-c(t-t_0)/2}\|\Omega(t_0)-\Omega_\infty\|_{HS}
  \]
        for all $t \ge t_0$.
\end{enumerate}
\end{theorem}

\begin{proof}
\emph{(1)} Both sides are invariant under the map $\Omega \mapsto \Theta \Omega \Theta^\dagger$, $\Theta \in K$, and every $\Omega$ near $C_*$ is $\Theta e^{R} D_* e^{-R} \Theta^\dagger$, where $R\in\mathfrak u(L)$ is skew-Hermitian and $R_{mn} \neq 0$ only for $\eta_{mn} > 0$. By \cref{lem:normal-hessian}, $\|[\Omega,\hat H]\|_{HS}^2 \ge 2\mu_* \sum \eta_{mn}\lvert R_{mn}\rvert^2 + O(\|R\|^3) = 2\mu_* ( V(\Omega) - V_* ) + O(\|R\|^3)$, and the cubic terms are absorbed for $c < 2\mu_*$ on a small neighborhood, which can be taken invariant by \cref{thm:almost-global} (2).

\emph{(2)} By \cref{lem:db-gradient} and item~(1), $\frac{d}{dt}\bigl(V(\Omega(t)) - V_*\bigr) = -\|[\Omega(t),\hat H]\|_{HS}^2 \le -c\,\bigl(V(\Omega(t)) - V_*\bigr)$ in $\mathcal U$, and the claim follows by Grönwall's inequality.

\emph{(3)} By \cref{thm:almost-global} (1) and item~(2), $\Omega(t)$ converges to a point $\Omega_\infty\in C_*$. Let $h(t):=\bigl(V(\Omega(t))-V_*\bigr)^{1/2}$. By \cref{lem:db-gradient} and item~(1), $\dot h=-\|[\Omega,\hat H]\|_{HS}^2/(2h)\le-\tfrac{\sqrt c}{2}\|[\Omega,\hat H]\|_{HS}$, and since $\|\dot \Omega\|_{HS}\le2\|\bar\Omega\|\,\|[\Omega,\hat H]\|_{HS}$, integrating from $t$ to $\infty$ and using item~(2) gives
\[
  \|\Omega(t)-\Omega_\infty\|_{HS}\le\int_t^\infty\|\dot \Omega(s)\|_{HS}\,ds
  \le\tfrac{4\|\bar\Omega\|}{\sqrt c}\,h(t)
  \le\tfrac{4\|\bar\Omega\|}{\sqrt c}\,e^{-c(t-t_0)/2}h(t_0).
\]
By \cref{thm:mb-lemma-global}, $V-V_*\le b\operatorname{dist}(\cdot,C_*)^2$ near $C_*$ for some $b>0$, so after shrinking $\mathcal U$, $h(t_0)\le\sqrt b\,\|\Omega(t_0)-\Omega_\infty\|_{HS}$. This gives (3) with $\gamma=4\|\bar\Omega\|\sqrt{b/c}$.
\end{proof}

\section{Convergence of the Iteration}
\label{app:descent}
In this appendix, we prove the results on the discrete iteration stated in Sec.~\ref{sec:iteration}.
\begin{proof}[Proof of Proposition~\ref{prop:descent}]
We prove~\eqref{eq:descent}, which states that, for $0<s_k<1/(2w_0\|\hat H\|)$,
\begin{equation*}
    V(\Omega_{k+1})\le V(\Omega_k)-s_k\big(1-2s_kw_0\|\hat H\|\big)\Big\|\comm{\Omega_k}{\hat H}\Big\|_{HS}^2 ,
\end{equation*}
where $V(\Omega)=\Tr(\Omega\hat H)$ is the weighted energy. Recall that the largest weight $w_0$ is the largest eigenvalue of $\Omega$.

Let $W:=W_k=\comm{\Omega_k}{\hat H}$ be the generator~\eqref{eq:Wk} of step $k$, which is skew-Hermitian, and $\Omega(t)=e^{tW}\Omega_ke^{-tW}$ for $0\le t\le s_k$. Since $\Omega_{k+1}=\Omega(s_k)$, we have every state updated as $\ket*{\phi^{(q)}_{k+1}}=e^{s_kW}\ket*{\phi^{(q)}_k}$ by~\eqref{eq:ket-update}. Let $V(t):=V(\Omega(t))=\Tr\big(\Omega(t)\hat H\big)$, so that $V(0)=V(\Omega_k)$ and $V(s_k)=V(\Omega_{k+1})$. Since $W$ is fixed during the step, $d\Omega(t)/dt=\comm{W}{\Omega(t)}$. Hence
\begin{equation*}
    V'(0)=\Tr\big(\comm{W}{\Omega_k}\hat H\big)=\Tr\big(W\comm{\Omega_k}{\hat H}\big)=\Tr(W^2)=-\|W\|_{HS}^2 ,
\end{equation*}
and
\begin{equation*}
    V''(t)=\Tr\big(\comm{W}{\comm{W}{\Omega(t)}}\hat H\big)=-\Tr\big(\comm{W}{\Omega(t)}\comm{W}{\hat H}\big).
\end{equation*}
Using the Cauchy--Schwarz inequality $|\Tr(XY)|\le\|X\|_{HS}\|Y\|_{HS}$, together with the inequality $\|\comm{X}{Y}\|_{HS}\le2\|X\|\,\|Y\|_{HS}$ and $\|\Omega(t)\|=w_0$, which holds because conjugation preserves the spectrum of $\Omega$, we obtain
\begin{equation*}
    |V''(t)|\le\|\comm{W}{\Omega(t)}\|_{HS}\,\cdot\Big\|\comm{W}{\hat H}\Big\|_{HS}\le4w_0\|\hat H\|\,\|W\|_{HS}^2 .
\end{equation*}
By Taylor's theorem, $V(s_k)=V(0)+s_kV'(0)+\tfrac12s_k^2V''(\xi)$ for some $\xi\in[0,s_k]$, so
\begin{equation*}
    V(s_k)\le V(0)-s_k\|W\|_{HS}^2+2s_k^2w_0\|\hat H\|\,\|W\|_{HS}^2 ,
\end{equation*}
which is~\eqref{eq:descent}. Under the step condition, the factor $s_k(1-2s_kw_0\|\hat H\|)$ is positive, so the weighted energy decreases strictly unless $W=\comm{\Omega_k}{\hat H}=0$.
\end{proof}

Combining Proposition~\ref{prop:descent} with Theorem~\ref{thm:flow} shows that, for almost every initial ensemble, the iteration also converges to $C_*$ for small step sizes. The proof only uses these two results and the fact that, for small steps, the iteration stays close to the flow over any finite time, so we only sketch it.

\begin{proof}[Proof sketch of Corollary~\ref{cor:iteration}]
By Appendix~\ref{app:morse-bott}, the critical points of $V$ form finitely many critical submanifolds $C_*,C_1,\dots,C_m$, and $V$ is constant on each of them. Recall that $V_*=\sum_{q=0}^pw_qE_q$ is its value on $C_*$, the minimum of $V$, and let $V_1:=\min_{j\ge1}V(C_j)>V_*$. The sublevel set $\mathcal S:=\{\Omega:V(\Omega)<V_1\}$ contains no critical points other than those in $C_*$.

\emph{(i) The flow enters $\mathcal S$ in finite time.} By Theorem~\ref{thm:flow}, for almost every $\Omega_0$ the flow converges to $C_*$, so $V(\Omega(T))<V_1$ at some finite time $T:=T(\Omega_0)$.

\emph{(ii) The iteration follows the flow up to time $T$.}
The iteration is a first-order discretization of the flow on a compact set, so on $[0,T]$ it stays within $\mathcal O(s)$ of the flow trajectory. For small enough $s$, it therefore also enters $\mathcal S$, at some step $k_0$.

\emph{(iii) In the set $\mathcal S$, the iteration converges to $C_*$.} Let $s<1/(2w_0\|\hat H\|)$. By Proposition~\ref{prop:descent}, $V(\Omega_k)\le V(\Omega_{k_0})<V_1$ for all $k\ge k_0$, so the iterates remain in $\mathcal S$. Since $V$ is bounded below by $V_*$, summing~\eqref{eq:descent} over all $k\ge k_0$ shows that $\sum_{k\ge k_0}\|\comm{\Omega_k}{\hat H}\|_{HS}^2$ is finite, so $\|\comm{\Omega_k}{\hat H}\|_{HS}\to0$. Hence the iterates approach critical points with $V<V_1$, which all lie in $C_*$, so $\operatorname{dist}(\Omega_k,C_*)\to0$.
\end{proof}

\begin{proof}[Proof of Proposition~\ref{prop:local-rate}]
Let $0<c<2\mu_*$ be as in the proposition. By item~(1) of Theorem~\ref{thm:local-rate-general}, there is an open neighborhood $\mathcal U$ of $C_*$ on which
\begin{equation*}
    \|\comm{\Omega}{\hat H}\|_{HS}^2\ge c\,\big(V(\Omega)-V_*\big).
\end{equation*}
Outside $\mathcal U$, we have $V>V_*$, since $V=V_*$ only on $C_*$. Because the set of ensembles outside $\mathcal U$ is closed and bounded and $V$ is continuous, $V-V_*$ has a positive minimum $\varepsilon$ there. Hence $V(\Omega)-V_*<\varepsilon$ implies $\Omega\in\mathcal U$. Since $V(\Omega_{k_0})-V_*<\varepsilon$ and $V$ does not increase by Proposition~\ref{prop:descent}, we have $\Omega_k\in\mathcal U$ for all $k\ge k_0$. For these $k$, the descent bound of Proposition~\ref{prop:descent} and the inequality above give
\begin{equation*}
    V(\Omega_{k+1})\le V(\Omega_k)-s(1-2sw_0\|\hat H\|)\|\comm{\Omega_k}{\hat H}\|_{HS}^2\le V(\Omega_k)-c\,s(1-2sw_0\|\hat H\|)\big(V(\Omega_k)-V_*\big),
\end{equation*}
and subtracting $V_*$ from both sides gives
\begin{equation*}
    V(\Omega_{k+1})-V_*\le\big(1-c\,s(1-2sw_0\|\hat H\|)\big)\big(V(\Omega_k)-V_*\big).
\end{equation*}
\end{proof}

\section{Implementation of the Weighted Phase Gate}
\label{app:phase-gate}
This appendix shows how to implement the weighted phase gate $e^{i\sqrt{s_k}D}$. As for the multi-controlled phase gate of DB-QITE (Ref.~\cite{gluza2026double}, App.~V.A), we compute a condition on the basis state into an ancilla, apply a phase to the ancilla, and uncompute it. Unlike in DB-QITE, $p+1$ inputs receive a phase, possibly different ones, and we use one such condition for each group of inputs that shares a phase. We consider $N$ qubits, so $L=2^N$. Each phase is applied to the ancilla with the single-qubit phase gate $P(\varphi):=\ket0\bra0+e^{i\varphi}\ket1\bra1$. To simplify notation, we omit the identity acting on the remaining qubits. For example, for a qubit $a$,
\begin{equation*}
    P(\varphi)_a := I\otimes\cdots\otimes I\otimes \underbrace{P(\varphi)}_{\text{qubit }a}\otimes I\otimes\cdots\otimes I .
\end{equation*}
We compare the constructions by the number of Toffoli gates used to compute the ancillas and the number of phase gates $P(\varphi)$.

\subsection{Inputs and Registers}
The gate $e^{i\sqrt{s_k}D}$, with $D=\sum_{q=0}^p w_q\ket{x_q}\bra{x_q}$ from Eq.~\eqref{eq:D}, multiplies each input $\ket{x_q}$, $q=0,\dots,p$, by $e^{i\theta_q}$, with $\theta_q:=\sqrt{s_k}\,w_q$, and acts as the identity on all other basis states. Without loss of generality, we take the inputs to be the first $p+1$ computational basis states, $\ket{x_q}=\ket q$. Here, for an integer $0\le x<2^N$, $\ket x$ denotes the $N$-qubit basis state whose bits are the binary representation of $x$. This is no restriction, since other inputs $\ket{x_q}=\Pi\ket q$, with $\Pi$ a permutation of basis states, are obtained by replacing the warm start $U_0$ of Sec.~\ref{sec:circuit} with $U_0\Pi$. Each $q\le p$ can be written with $M$ bits, where $M:=\lceil\log_2(p+1)\rceil$ is the number of bits of $p$. We call the last $M$ qubits the index register and the first $N-M$ qubits the top qubits. Every input then has its top qubits in $\ket0$ and holds $q$ in its index register. For an integer $a$, we write $a_j$ for the $j$th bit of its binary representation, with $a_0$ the least significant one. For a basis state $\ket x$, we write $y$ for the number held in its index register, so that $x=y$ when the top qubits are in $\ket0$. The gate is then
\begin{equation} \label{eq:dbmex-gate}
e^{i\sqrt{s_k}D}=I+\sum_{q=0}^{p}\big(e^{i\theta_q}-1\big)\ket q\bra q =\mathrm{diag}\big(e^{i\theta_0},\dots,e^{i\theta_p},1,\dots,1\big),
\end{equation}
which applies phases to the first $p+1$ basis states, i.e., those with $x\le p$.

A direct construction follows from the fact that the projectors $\ket q\bra q$ commute, so that the gate is the product $\prod_{q=0}^{p}e^{i\theta_q\ket q\bra q}$ of $p+1$ multi-controlled phase gates, one per input, each applying a phase to a single basis state. Implementing each factor with a ladder of $N-1$ Toffoli gates gives $(p+1)(N-1)$ Toffoli gates and $p+1$ phase gates, for any weights. The rest of this appendix gives a more efficient construction.

\subsection{Block Construction}
\begin{figure}[H]
\centering
\resizebox{\linewidth}{!}{
\begin{quantikz}[row sep=0.3cm, column sep=0.5cm]
\lstick[2]{$\ket\psi$} & \qwbundle{N-M\ \mathrm{top}} &  &  &  & \octrl{2}\gategroup[3,steps=1,style={dashed,rounded corners,inner xsep=4pt},label style={label position=above}]{$T$} &  &  &  &  &  &  &  &  &  & \octrl{2}\gategroup[3,steps=1,style={dashed,rounded corners,inner xsep=4pt},label style={label position=above}]{$T^\dagger$} &  & \rstick[2]{$e^{i\sqrt{s_k}D}\ket\psi$} \\
 & \qwbundle{M\ \mathrm{index}} &  &  &  &  &  & \gate[3]{G_1} &  & \gate[3]{G_1^\dagger} & \gate[3,style={draw=none,fill=none}]{\ \cdots\ } & \gate[3]{G_B} &  & \gate[3]{G_B^\dagger} &  &  &  & \\
\lstick{$\ket t=\ket0$} &  &  &  &  & \targ{} &  &  &  &  &  &  &  &  &  & \targ{} &  & \rstick{$\ket0$} \\
\lstick{$\ket m=\ket0$} &  &  &  &  &  &  &  & \gate{P(\varphi_1)} &  &  &  & \gate{P(\varphi_B)} &  &  &  &  & \rstick{$\ket0$}
\end{quantikz}
}
\caption{Weighted phase gate with $B$ blocks, Eq.~\eqref{eq:block-construction}. All ancillas start and end in $\ket0$. Work ancillas of $T$ and $G_b$ are not shown.}
\label{fig:blocks}
\end{figure}
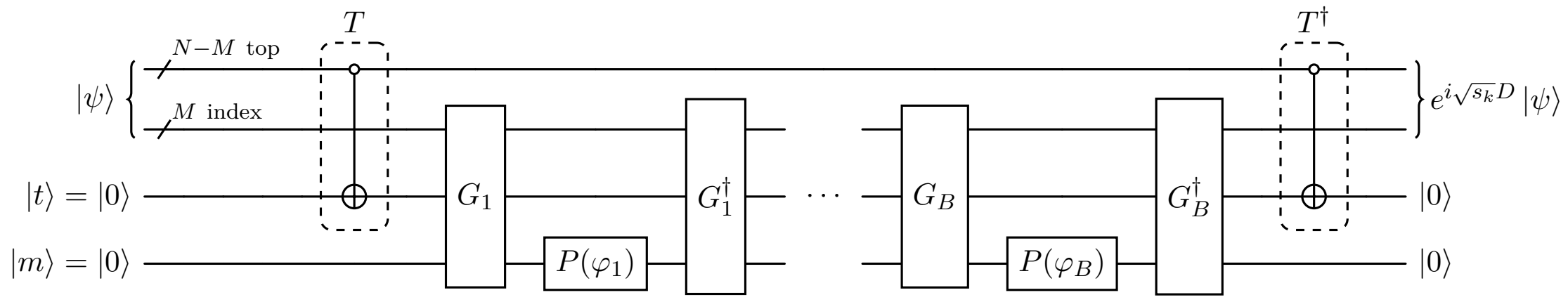

A basis state is an input if and only if its top qubits are in $\ket0$ and $y\le p$. The construction uses two dedicated ancillas and a few work ancillas for intermediate results, all initialized in $\ket0$. The first ancilla, $t$, records whether the top qubits are in $\ket0$. The second, the marker $m$, is set to $1$ exactly on the basis states that should receive a given phase, so that this phase can be applied by a single phase gate on $m$. A circuit $T$ computes
\begin{equation} \label{eq:t-def}
    t:=[\text{top qubits are in $\ket0$}]
\end{equation}
into the ancilla $t$, where $[C]$ denotes $1$ if the condition $C$ holds and $0$ otherwise. It uses a ladder of $N-M-1$ Toffoli gates, each combining the previous result with one more qubit. For $M=N$, there are no top qubits, $t\equiv1$, and $T$ and the ancilla $t$ are omitted. The phases are then applied in blocks. A block $b$ is given by a set $S_b\subseteq\{0,\dots,p\}$ of indices and an angle $\varphi_b$. Its circuit $G_b$ computes $m_b=t\wedge[y\in S_b]$ into the marker, using $t$ and the index register, $P(\varphi_b)$ acts on the marker, and $G_b^\dagger$ uncomputes it. Each block thus adds $\varphi_b$ to the angle of every input in $S_b$. We choose the blocks such that each input lies in exactly one block, whose angle is the angle $\theta_q$ of that input. Then
\begin{equation} \label{eq:block-construction}
    e^{i\sqrt{s_k}D}=T^\dagger\Big[\prod_b G_b^\dagger\,P(\varphi_b)_m\,G_b\Big]\,T ,
\end{equation}
as shown in Fig.~\ref{fig:blocks}. On a basis state $\ket x$, block $b$ writes $m_b$ into the marker, applies $e^{i\varphi_b}$ if $m_b=1$, and uncomputes the marker. Since $P(\varphi_b)$ is diagonal, it changes no bit, so $G_b^\dagger$ returns the marker to $\ket0$. Each block thus acts on the system as a diagonal gate, and the blocks commute. A non-input has $t=0$ or $y>p$, so all its markers are $0$, and it receives no phase. Input $q$ lies in exactly one block and receives its angle $\theta_q$. Finally, $T^\dagger$ uncomputes $t$. The weight choices below differ only in the blocks they use.

\subsection{Single-Index and Comparator Blocks}
To cover the different weight choices of Sec.~\ref{sec:weights}, two kinds of blocks suffice, namely single-index blocks and comparator blocks. A single-index block has $S_b=\{q\}$, and its marker $m_q:=[x=q]=t\wedge[y=q]$ is the AND of $t$ and the index bits, with $y_j$ negated where $q_j=0$. It is computed with a ladder of at most $M$ Toffoli gates. A comparator block has $S_b=\{0,\dots,r\}$ for some $r\le p$, and its marker is $m_{\le r}:=[x\le r]=t\wedge[y\le r]$. We write $G_{\le r}$ for its circuit.

To compare $y$ with $r$, we scan the bits from the least significant one and keep a running bit $c_j$, which records whether $y\le r$ when only the lowest $j+1$ bits are taken into account,
\begin{equation*}
    c_j:=[(y_j\dots y_0)\le(r_j\dots r_0)],
\end{equation*}
with $c_{-1}:=1$. Each $c_j$ is computed from $c_{j-1}$ and the single new bit $y_j$, and written into a work ancilla,
\begin{equation} \label{eq:comparator}
    c_j=\begin{cases}\bar y_j\wedge c_{j-1}, & r_j=0,\\ \bar y_j\vee c_{j-1}, & r_j=1,\end{cases}
\end{equation}
since two numbers are ordered by the highest bit in which they differ. If $y_j\ne r_j$, bit $j$ decides the comparison, and if $y_j=r_j$, then $c_j=c_{j-1}$. Then $c_{M-1}=[y\le r]$. Each step costs one Toffoli gate together with $X$ gates, using $\bar y_j\vee c_{j-1}=\overline{y_j\wedge\bar c_{j-1}}$ for $r_j=1$. A final Toffoli gate writes $m_{\le r}=t\wedge c_{M-1}$ into the marker.

If $r_{j-1}=\dots=r_0=1$, then $(y_{j-1}\dots y_0)\le(r_{j-1}\dots r_0)$ holds for every $y$, so $c_{j-1}\equiv1$. Step $j$ then needs no gate, because $c_j$ is $1$ if $r_j=1$, and $\bar y_j$ if $r_j=0$, which is read directly from $y_j$. The steps are free up to and including the lowest $0$ of $r$, and each later step costs one Toffoli gate. The comparator therefore takes at most $M-1$ Toffoli gates and the final step one more, so that, with the $N-M-1$ Toffoli gates of $T$, the marker $m_{\le r}$ takes at most $N-1$ Toffoli gates. If $r=2^M-1$, all bits of $r$ are $1$, $c_{M-1}\equiv1$, and $m_{\le r}=t$. If $M=N$, $T$ and the final step are absent, and the bound still holds.

\subsection{Weight Choices and Cost}
We now give the gate construction for the three weight choices of Sec.~\ref{sec:weights}. 
\begin{itemize}
    \item \textit{Equal weights}, $w_q=w_0$. We use one comparator block on $\{0,\dots,p\}$ with angle $\theta:=\sqrt{s_k}\,w_0$, which gives every input the angle $\theta$. Then
    \begin{equation} \label{eq:form-equal}
    e^{i\sqrt{s_k}D}=T^\dagger\big[G_{\le p}^\dagger\,P(\theta)_m\,G_{\le p}\big]\,T ,
\end{equation}
where $G_{\le p}$ computes $m_{\le p}=[x\le p]$ into the marker. The gate costs at most $N-1$ Toffoli gates and one phase gate. If $p+1=2^M<2^N$, all bits of $p$ are $1$, so every index satisfies $y\le p$ and $m_{\le p}=t$. The comparator and the marker are then not needed, and the phase gate acts on $t$ directly, so that Eq.~\eqref{eq:form-equal} becomes the multi-controlled phase gate on the top qubits, $T^\dagger P(\theta)_tT=e^{i\theta\ket0\bra0^{\otimes(N-M)}}$, with $N-M-1$ Toffoli gates.

\item \textit{Targeted weights}, $w_0=\dots=w_{p-1}$ and $w_p=w$, with $p \geq 1$. Let $\theta:=\sqrt{s_k}\,w_0$ and $\theta_p:=\sqrt{s_k}\,w$. We use two blocks, a comparator block on $\{0,\dots,p-1\}$ with angle $\theta$ and a single-index block for $q=p$ with angle $\theta_p$. Inputs $0,\dots,p-1$ then receive $\theta$, and input $p$ receives $\theta_p$. Thus
\begin{equation} \label{eq:form-targeted}
    e^{i\sqrt{s_k}D}=T^\dagger\,\big[G_{\le p-1}^\dagger\,P(\theta)_m\,G_{\le p-1}\big]\big[G_p^\dagger\,P(\theta_p)_m\,G_p\big]\,T ,
\end{equation}
where $G_{\le p-1}$ computes $m_{\le p-1}=[x\le p-1]$ and $G_p$ computes $m_p=[x=p]$ into the marker. The gate costs at most $N+M-1$ Toffoli gates and two phase gates.

\item \textit{Distinct weights}, with arbitrary $\theta_0,\dots,\theta_p$. We use $p+1$ blocks, one single-index block for each input $q$ with angle $\theta_q$, so each input receives its own angle, and
\begin{equation} \label{eq:form-general}
    e^{i\sqrt{s_k}D}=T^\dagger\Big[\prod_{q=0}^{p}G_q^\dagger\,P(\theta_q)_m\,G_q\Big]\,T ,
\end{equation}
where $G_q$ computes $m_q=[x=q]$ into the marker. The gate costs $p+1$ phase gates and at most $N-M-1+(p+1)M$ Toffoli gates.

\end{itemize}

Table~\ref{tab:gate-cost} summarizes the cost of the weighted phase gate for inputs $\ket q$, along with that of the multi-controlled phase gate in DB-QITE~\cite{gluza2026double}.

\begin{table}[H]
\centering
\small
\renewcommand{\arraystretch}{1.25}
\begin{tabular}{@{}lllll@{}}
\toprule
Algorithm & Weights & Blocks & Toffoli gates & Phase gates \\
\midrule
DB-QITE & $w_0=1$ & none & $N-1$ & $1$ \\
\midrule
DB-MEX & equal & $1$ comparator & $\le N-1$ & $1$ \\
 & targeted & $1$ comparator, $1$ single & $\le N+M-1$ & $2$ \\
 & distinct & $p+1$ single & $\le N-M-1+(p+1)M$ & $p+1$ \\
\bottomrule
\end{tabular}
\caption{Cost of one weighted phase gate on $N$ qubits, with the inputs $\ket q$ for DB-MEX and ${M=\lceil\log_2(p+1)\rceil}$. Toffoli gates are counted for the compute step only.}
\label{tab:gate-cost}
\end{table}

For equal weights, the weighted phase gate costs no more than the multi-controlled phase gate of DB-QITE. Targeted weights add at most $M$ Toffoli gates and one phase gate, and distinct weights add at most $M$ Toffoli gates and one phase gate per additional state. Hence, for the small $p$ of interest, the weighted phase gate costs about the same as that of DB-QITE.

\end{document}